\documentclass[12pt,a4paper]{scrartcl}%
\usepackage[latin1]{inputenc}
\pdfoutput=1 
\usepackage{setspace}
\usepackage{amsmath}
\usepackage{amsfonts}
\usepackage{amssymb}
\usepackage{nicefrac}
\usepackage{siunitx}%
\usepackage{hyperref}
\usepackage{transparent}
\usepackage[numbers]{natbib}
\usepackage{wrapfig}
\usepackage{subcaption}
\usepackage{longtable}
\usepackage{booktabs}
\usepackage[T1]{fontenc}
\usepackage{import}
\usepackage{comment}
\usepackage{pgfplots}
\usepackage{pgfplotstable}
\usepgfplotslibrary{fillbetween}
\pgfplotsset{
  log x ticks with fixed point/.style={
      xticklabel={
        \pgfkeys{/pgf/fpu=true}
        \pgfmathparse{exp(\tick)}%
        \pgfmathprintnumber[fixed relative, precision=3]{\pgfmathresult}
        \pgfkeys{/pgf/fpu=false}
      }
  },
  log y ticks with fixed point/.style={
      yticklabel={
        \pgfkeys{/pgf/fpu=true}
        \pgfmathparse{exp(\tick)}%
        \pgfmathprintnumber[fixed relative, precision=3]{\pgfmathresult}
        \pgfkeys{/pgf/fpu=false}
      }
  }
}
\pgfplotsset{compat=1.3}
\usepgfplotslibrary{external} 
\usetikzlibrary{arrows.meta,shapes,fit,backgrounds}
\tikzset{external/system call={pdflatex \tikzexternalcheckshellescape -halt-on-error
-interaction=batchmode -jobname "\image" "\texsource"}}

\author{T. Antritter\textsuperscript{a,1}, T. Josyula\textsuperscript{a}, T. Mari\'c\textsuperscript{b}, D. Bothe\textsuperscript{b}, P. Hachmann\textsuperscript{c},\\ B. Buck\textsuperscript{c}, T. Gambaryan-Roisman\textsuperscript{a}, P. Stephan\textsuperscript{a}}

\title{A Two-Field Formulation for Surfactant Transport within the Algebraic Volume of Fluid Method}
\begin{document}
\maketitle

\noindent\textsuperscript{a} Institute for Technical Thermodynamics, Technische Universit\"at Darmstadt, Germany \\
\textsuperscript{b} Institute for Mathematical Modeling and Analysis, Technische Universit\"at Darmstadt, Germany \\
\textsuperscript{c} Heidelberger Druckmaschinen AG, Wiesloch, Germany \\
\textsuperscript{1} T.A.'s present address: BASF SE, Ludwigshafen am Rhein, Germany \\

\abstract{
\noindent 
\textbf{Abstract}

Surfactant transport plays an important role in many technical processes and industrial applications such as chemical reactors, microfluidics, printing and coating technology. High fidelity numerical simulations of two-phase flow phenomena reveal rich insights into the flow dynamics, heat, mass and species transport. In the present study, a two-field formulation for surfactant transport within the algebraic volume of fluid method is presented.
The slight diffuse nature of representing the interface in the algebraic volume of fluid method is utilized to track the concentration of surfactant at the interface as a volumetric concentration.
Transport of insoluble and soluble surfactants is investigated by tracking two different concentrations of the surfactant, one within the bulk of the liquid and the other one at the interface. These two transport equations are in turn coupled by source terms considering the ad-/desorption processes at a liquid-gas interface.
Appropriate boundary conditions at a solid-fluid interface are formulated to ensure surfactant conservation, while also enabling to study the ad-/desorption processes at a solid-fluid interface. The developed numerical method is verified by comparing the numerical simulations with well-known analytical and numerical reference solutions. The presented numerical methodology offers a seamless integration of surfactant transport into the algebraic volume of fluid method, where the latter has many advantages such as volume conservation and an inherent ability of handling large interface deformations and topological changes.
\\
\\
\textbf{Author contributions}\\
T.A.: Methodology, Software, Validation, Visualization, Writing -
original draft;\\
T.J.: Software, Visualization, Validation, Writing -
original draft;\\
T.M.: Software, Data curation, Writing - review \& editing;\\
D.B.: Funding acquisition, Validation, Writing - review \& editing;\\
P.H.: Conceptualization, Methodology, Writing - review \& editing;\\
B.B.: Conceptualization, Funding acquisition, Supervision, Writing - review \& editing;\\
T.G.-R.: Funding acquisition, Methodology, Project administration, Supervision, Writing - review \& editing;\\
P.S. Funding acquisition, Methodology, Project administration, Supervision, Writing - review \& editing
}\\

\section{Introduction}

Surface active substances, so called surfactants, play an important role in many technical and natural free-surface or multiphase flow phenomena. Surfactants accumulate at liquid-liquid or liquid-gas interface, where they lower the surface tension. Different concentrations along the interface may therefore cause surface tension gradients, further leading to solutal Marangoni-flow. In technical applications, surfactants may be deliberately added to adjust the surface tension of the liquid. In agricultural applications, surfactants are used as adjuvants, to improve wetting and the uptake of the pesticide by plants \citep{Jibrin2021}.%
In the chemical industry, surfactants can play a role, e.g. in bubble-column reactors, where components with lower surface tension within the liquid mixture may accumulate at the bubble interfaces and the resulting Marangoni flow due to the concentration gradients can immobilize the liquid-gas interface \citep{frumkin1947surfactants,clift2005bubbles}.  This accumulation can substantially change aspects of the flow, such as e.g. decreasing the bubble rise velocity \citep{Alves2005effect}. Further, surfactants are shown to significantly effect the heat transfer in boiling studies, where the heat flux can either increase or decrease depending on the surfactant type and concentration \citep{mata2022dynamic}. In inkjet printing, surfactants are added to ink formulations in order to facilitate drop generation from the print head as well as to adjust the wetting and deposition behavior of the ink on the print substrate \citep{Graindourze2018,Mulla2016}. Interestingly, in inkjet printing, the presence of a surfactant can also help minimize the well-known artifact known as "coffee-ring" \citep{still2012surfactant}.  

Numerical simulation allows a detailed analysis of complex flow phenomena. A variety of  methods for the simulation of free surface and multiphase flows have been developed \citep{prosperetti-tryggvason-2007, tryggvason-scardovelli-zaleski-2011, Woerner2012}. According to \citet{Woerner2012}, these methods can be grouped by their representation of the liquid-fluid interface. The sharp interface methods, such as the arbitrary Lagrangian-Eulerian (ALE) method, the front-tracking (FT) approach as well as the geometric volume of fluid method (VOF) and the level-set method (LS) represent the interface with zero thickness. Opposed to that, the phase-field method and the algebraic volume of fluid method represent the interface with a finite thickness of a few mesh spacings. 

Furthermore, methods may be grouped into Lagrangian and Eulerian descriptions \citep[cf.][]{Woerner2012}. In Lagrangian methods, the interface is represented by discrete points on the interface, which are tracked with the flow. The front-tracking method \citep{Unverdi1992} follows such an approach in order to track the interface, while pressure and velocity are calculated on an Eulerian background mesh. The surface mesh can furthermore be used to discretize a transport equation for surfactants on the interface. Such an approach is presented by \citet{Muradoglu2008}, and is demonstrated for expanding drops, rising drops due to surface tension gradients and buoyancy, and drops splitting under the influence of surfactant. The ALE method \citep{Hirt1974}, which allows blending between Lagrangian and Eulerian formulations, can similarly take advantage of a discrete representation of the interface by ensuring mesh motion with the flow in interface normal direction. An ALE based interface tracking  method in the context of the unstructured finite-volume method is presented by \citet{Tukovic2012}. Building on \citep{Tukovic2012}, \citet{Dieter-Kissling2015a} developed an ALE approach approach including surfactants by using the finite area method to describe surfactant transport on the interface. They furthermore applied it to drops covered with insoluble surfactant \citep{Dieter-Kissling2015a}, as well as drops with soluble surfactant \citep{Dieter-Kissling2015b}. Later on, \citet{Pesci2018} extended the method by a subgrid-scale model to account for species transfer processes at the interface and applied it to gas bubbles rising under the influence of a soluble surfactant. While the representation of the interface with a surface mesh is obviously beneficial when describing transport phenomena on the interface, such as surfactant transport, the surface mesh representation in FT and ALE methods requires special care when topological  changes, such as drop impact on a substrate, or coalescence occur. A detailed description of the necessary steps to capture such topological changes is given by \citet{Tryggvason2001} for their front-tracking approach.

In Eulerian methods, the phase distribution within the computational domain is described on a background mesh using a scalar transport equation. In the LS method and conservative level-set (C-LS) \citep{Olsson2005} method, the scalar is a signed distance to the interface and in PF methods, the phase-field function. In case of the VOF method, this scalar is the volume fraction of one of the phases.
The VOF method can further be classified based on the way, the advection of the phase indicator function is numerically treated.
In the geometric VOF method, phase-specific volumes fluxed across cell faces are geometrically computed using the flow velocity and the reconstructed phase indicator. Since the advection of volume fractions associated with the background Eulerian mesh uses Lagrangian tracking, the geometric VOF method is a Lagrangian-Eulerian method. A review of un-split geometric VOF methods can be found in \citet{maric2020unstructured}. On the other hand, in the algebraic VOF method, the fluxes are calculated using the volume fraction at the cell faces estimated by algebraic interpolation of the cell centered values, using high-resolution differencing schemes \citep{UBBINK1999,Xiao2005}. The algebraic VOF solver \emph{interFoam} of the open-source CFD library OpenFOAM is a widely used implementation of this approach \citep{Deshpande2012}, which is also used throughout the work presented in this article. Due to the description of phase distribution on a background mesh using a scalar transport equation, the above mentioned Eulerian methods are inherently able to handle strong deformations of the interface or topological changes without a mesh update. Nevertheless, local mesh refinement at the interface will still be beneficial to maintain a high resolution at the interface in order to capture the relevant local physical phenomena. A software-framework extending the open-source CFD library OpenFOAM \citep{OpenFOAM2016} using adaptive mesh refinement including load balancing during parallel execution of the code is presented  by \citet{Rettenmaier2019}, following works of \citet{Voskuilen2014} and \citet{Batzdorf2015}. This mesh refinement is especially relevant for coalescence, as in numerical simulations, drops coalesce once their distance can no longer be resolved by the numerical mesh. Without additional measures, this results in a mesh dependence of the so-called numerical coalescence \citep{Woerner2012}. Furthermore, the overlapping interface that may occur in methods with finite interface results in drops coalescing more easily \citep{Woerner2012}.

Despite these limitations, the above mentioned benefits regarding an efficient handling of topological changes led to a wide usage of the algebraic VOF method across different engineering applications. It is utilized in various multiphase fluid flow problems such as rising bubbles due to due to buoyancy \citep{Deising2018}, flow in microchannels \citep{Hoang2013}, microfluidic T-junctions \citep{Nekouei2017}, and flow in porous media \citep{Raeini2012}. Furthermore, the VOF method is also used across different scales, ranging from small scale boiling studies in microgravity \citep{Franz2021paper}, to marine applications \citep{Higuera2013}. The method has also been proven to handle contact line movement observed in various wetting scenarios \citep{Franz2021paper,Batzdorf2017,Berberovic2011,Kunkelmann2009,Sikalo2005}.

The lack of a discrete representation of the interface in the above-mentioned Eulerian methods in general, and the algebraic VOF method in particular, however, poses a challenge when dealing with interfacial transport phenomena. Nevertheless, several approaches have been presented in the past. For the geometric VOF method, \citet{James2004}, as well as \citet{Alke2009} made use of the geometric reconstruction of the interface for the advection of the volume fraction to also advect surface active substances with and along the interface. \citet{Lakshmanan2010} made use of the finite thickness of the interface in the C-LS method and formulated a transport equation for the surfactant, which ensures transport with and within this interface region. Later, \citet{CLERETDELANGAVANT2017271} presented a methodology to investigate the transport of soluble surfactants using the LS method. For the phase field method, \citeauthor{Teigen2009} \citep{Teigen2009,Teigen2011} introduced a method, which is based on the concept of constant normal extension. Some background information for this approach can be found in the work of \citet{Cermelli2005}. The general concept of the method is to describe species transport with and along the interface by solving a transport equation for the local surface excess concentration, but extended to a neighborhood of the interface using values in interface normal direction. Transport within the bulk phase is described by a separate balance equation. Solubility of the surfactant within the bulk phase is  modeled by coupling the transport of two fields. The Cahn-Hilliard equation based phase-field method offers an alternative approach. Similar to describing the transport of the evolution of the interface using a chemical potential as function of the phase-field function, \citet{Soligo2019} described transport of surfactant along the interface and within the bulk using a Cahn-Hilliard-like equation.

In the present work, a method to incorporate surface active substances into the framework of the algebraic VOF method is presented. Allowing to account for surfactant transport will further improve the versatility of the method. Building on the available open-source software OpenFOAM \citep{OpenFOAM2016} and additional developments regarding wetting \citep{Berberovic2011,Sikalo2005} and coupled wetting and heat transfer phenomena \citep{Kunkelmann2009,Batzdorf2017,Franz2021paper}, this will allow investigation of challenging multiphysics phenomena including surfactant transport. Such multiphysics phenomena are relevant, e.g., in inkjet printing, where a  surfactant-laden ink \citep{Mulla2016} is ejected from a heated print head \citep[see e.g.][]{Corrall2018} onto the substrate. In our work, a two-field formulation for the surfactant concentration is used, where the surfactant concentration at the interface and within the bulk are tracked separately. Specifically, the surface excess concentration at a liquid-gas interface is tracked using a volumetric interfacial excess concentration, taking advantage of the slightly diffuse interface region. Interestingly, this approach utilizing the diffuse nature of the interface region was employed before in the context of C-LS \citep{Lakshmanan2010} and phase field methods \citep{Soligo2019}, but not within the algebraic VOF method. Moreover, by considering the solubility of surfactant within the bulk, the numerical framework can handle the adsorption and desorption at liquid-gas and solid-liquid interfaces. Within our proposed method, this is handled by coupling the transport equations for the bulk and the interface. The numerical methodology presented here is a further development of the method introduced in \citet{Antritter2019b}, where it was used to study the spreading of micrometer sized drops, relevant to inkjet printing \citep{Antritter2019b}. Therein, the main focus was on the effect of solubility of a surfactant on the overall spreading behaviour of a drop, where the treatment of a moving contact line is crucial \citep{Antritter2019b}. In the present work, the numerical methodology and the implementation is also extended for considering the adsorption at a wall and the developed numerical method is verified rigorously with a wide range of verification cases. 

The remainder of the paper is organized as follows. In Section \ref{sec:Method Description}, the governing equations and numerical methodology are presented where, the algebraic VOF method is introduced first, followed by the two-field formulation for surfactant transport. Finally, numerical aspects and the solution procedure are discussed. In Section \ref{sec:Verification}, the verification of the numerical model is presented. First, the transport of an insoluble surfactant is considered, i.e. the case in which the surfactant is present only at a liquid-gas interface. Then, the transport of a soluble surfactant is presented, where adsorption at a liquid-gas interface and at a solid-liquid interface are considered, respectively. Finally, a verification case for Marangoni flow generated along the interface of a drop is presented. In Section \ref{sec:Conclusion}, the conclusion of the present work is presented, followed by a brief outlook towards future work.

\section{Governing Equations and Numerical Methodology}

\label{sec:Method Description}
This section first presents the governing equations for the algebraic volume of fluid method, followed by the introduction of our two-field approach for surfactant transport. Special emphasis is put on the coupling of the bulk and interface concentration fields. Finally, the discretization schemes and solution procedure are discussed. The present method is a further development of the method introduced and employed in \citet{Antritter2019b}.

\subsection{Algebraic VOF}
Within the Volume of Fluid method, introduced by \citet{Hirt1981}, the distribution of the phases within the computational domain is described by the volume fraction field, representing the fractional volume of one of the two phases within a control volume, here denoted by $\alpha$. In the present work, the scalar $\alpha$ refers to the volume fraction of the liquid. Accordingly, an $\alpha$-value of 1 indicates that the control volume is fully occupied by the liquid and an $\alpha$-value of 0 indicates that the control volume is fully occupied by the gas phase. Consequently, $0<\alpha<1$ indicates the presence of the liquid-gas interface. Within the algebraic VOF method, transport of the phases is implicitly described by the advection equation
\begin{equation}
	\label{eq:transportEqn}
	\frac{\partial \alpha}{\partial\tau} +  \nabla\cdot(\alpha \mathbf{u}) + \nabla\cdot(\alpha(1-\alpha)\mathbf{u}_\mathrm{r})= 0, 
\end{equation}
where $\tau$ represents time and $\mathbf{u}$ is the velocity. Due to the steep transition of the $\alpha$ field from $0$ to $1$ at the interface, special care has to be taken regarding its advection. In \autoref{eq:transportEqn}, $\mathbf{u}_\mathrm{r}= |\mathbf{u}\cdot\mathbf{n}_f| \mathbf{n}_\kappa$ is an artificial relative compressive velocity, acting in interface normal direction $\mathbf{n}_\kappa$, counteracting the smearing of the interface by numerical diffusion. Above, $\mathbf{n}_f$ denotes the cell face unit normal vector.
The above equation, together with the continuity equation for incompressible fluids
\begin{equation}
	\label{eq:continuityEqn}
	\nabla \cdot \mathbf{u} = 0,
\end{equation}
forms the mass balance for the two phases. Assuming Newtonian fluids, the momentum balance is given by
\begin{align}
	\label{eq:momentumBalance}
	\frac{\partial \rho_\mathrm{m} \mathbf{u}}{\partial \tau} &+ \nabla \cdot (\rho_\mathrm{m} \mathbf{u} \otimes \mathbf{u}) = \nabla \cdot \left[\mu_\mathrm{m} \left(\nabla \mathbf{u}  + (\nabla \mathbf{u})^\mathsf{T}\right)\right] -\nabla p + \rho_\mathrm{m} \mathbf{g} + \mathbf{f}_\mathrm{\sigma},
\end{align}%
where $\rho_\mathrm{m}$ and $\mu_\mathrm{m}$ are averaged density and viscosity of the phases. Throughout this work, averaged properties of any property $x$ are calculated according to $x_\mathrm{m}=\alpha\, x_\mathrm{\ell} + (1-\alpha) x_\mathrm{g}$. The indices $\mathrm{\ell}$ and $\mathrm{g}$ stand for liquid and gas phase, respectively. Pressure is denoted by $p$, and $\mathbf{g}$ is the acceleration due to gravity. The capillary forces are transformed into body forces according to
\begin{equation}
\label{eq:fc}
	\mathbf{f}_\sigma = \sigma_\mathrm{\ell g} \kappa \delta_2 \mathbf{n}_\sigma + \delta_2 \nabla_\Sigma \sigma_\mathrm{\ell g}
\end{equation}%
using the continuum surface force (CSF) approach introduced by \citet{Brackbill1992}.
Here, $\mathbf{f}_\mathrm{\sigma}$ accounts for the Laplace pressure jump across the interface as well as Marangoni tension in interface tangential direction. Therein, $\sigma_\mathrm{\ell g}$ is the surface tension of the liquid-gas interface and $\nabla_\Sigma=(\mathbb{I}-\mathbf{n}_\kappa\otimes\mathbf{n}_\kappa)\nabla$ is the interfacial gradient. If surfactants are present, the surface tension depends on their local surface excess concentration. The surface equation of state describing this dependence on the one hand, and the adsorption isotherm describing the equilibrium of bulk and surface excess concentrations on the other hand, are related through Gibb's adsorption equation. An example for Langmuir-Freundlich adsorption kinetics and the respective equation of state is given in \autoref{sec:AppendixFreundlich}.
The interface normal direction for the Laplace pressure contribution $\mathbf{n}_\sigma$ is evaluated from the volume fraction field as
\begin{equation}
\mathbf{n}_\sigma=\frac{\nabla\alpha}{||\nabla\alpha||_2}.
\end{equation}
In the present work, we employ $\delta_2$ as a scaled form of the regularized Dirac delta. Opposed to the original formulation of the CSF model, where $\delta_\mathrm{I} = ||\nabla\alpha||_2$ is used, here we follow the density-scaled CSF model proposed for the level-set method by \citet{Yokoi2014}. Transferring this approach to the VOF context, $\delta_{2}$ is then calculated as \citep{Antritter2019b}
\begin{equation}
\delta_2 = ||\nabla\alpha^2||_2 = 2\alpha||\nabla\alpha||_2
\end{equation}
and, hence,
\begin{equation}
\delta_2\mathbf{n}_\sigma=2 \alpha\,\nabla\alpha = \nabla \alpha^2.
\end{equation}
This approach shifts capillary forces towards the phase with $\alpha = 1$. Thus, with $\alpha$ representing the volume fraction of the higher density phase, accelerations due to capillary forces are reduced, as compared to the standard CSF model. This also results in a reduction in artificial velocities arising at the interface due to inaccuracies in the evaluation of capillary forces, so-called spurious or parasitic currents \citep{Yokoi2014}.

Furthermore, we calculate the interface normals according to
\begin{equation}
\label{eq:normalCalculation}
\mathbf{n}_\kappa = \frac{\nabla \alpha_\mathrm{S}}{||\nabla \alpha_\mathrm{S}||_2 + \varsigma_{\nabla\alpha}}
\end{equation}
for the evaluation of interface curvature 
\begin{equation}
\label{eq:curvature}
\kappa = -\nabla \cdot \mathbf{n}_\kappa
\end{equation}
from a smoothed volume fraction field $\alpha_\mathrm{S}$. Therein, $\varsigma_{\nabla\alpha}$\footnote{Within the scope of this work, $\varsigma_{\nabla\alpha}$ = $10^{-8}/\sqrt[3]{V_\mathrm{avrg}}$ is used, where $V_\mathrm{avrg}$ is the average cell volume of the domain. Compared to $ ||\nabla \alpha_\mathrm{S}||_2$, this value is negligible  within the interface region.} is a small value preventing division by zero outside the interface region. In order to obtain the smoothed volume fraction field, the diffusion equation 
\begin{equation}
\label{eq:diffSmoothingAlpha}
\frac{\partial \alpha_\mathrm{S}}{\partial{\tau^*}} = D_\mathrm{S} \nabla^2 \alpha_\mathrm{S}
\end{equation}
is integrated in pseudo time $\tau^*$ up to $\tau^*=\tau^*_\mathrm{S}$, starting with $\alpha_{S,0}=\alpha$ at $\tau^*=0$. The overall degree of smoothing is, therefore, defined by the product $D_\mathrm{S}\tau^*_\mathrm{S}$. For the verification cases presented in \autoref{sec:MarangoniFlow},
a smoothing parameter of $D_\mathrm{s}\tau^*_\mathrm{S}=\SI{5e-14}{\m\squared}$ is employed. This smoothing results in a reduction of curvature errors with short wavelength, increasing the accuracy of interface curvature for smoother interfaces. However, it should be noted that this approach will also systematically underestimate physical interface curvatures of short wavelength, if $D_\mathrm{s}\tau^*_\mathrm{S}$ is chosen too large for the considered problem. 

\subsection{Two-Field Approach for Surfactant Transport}
Surfactant transport within the fluid domain is described using a two-field approach, where one field describes surfactant dissolved within the liquid bulk and the second field represents the excess surfactant present at the liquid-gas interface.

\subsubsection{Bulk Concentration}
Transport of surfactant within the liquid bulk is described based on the continuum species transport (CST) model by \citet{Deising2016}. Using conditional volume averaging, they arrive at a single field transport model for species transport within bulk phases and across the phase interface. For the surfactant transport within the present model, the limiting case of this CST model for solubility in only one of the two phases is considered.
Thus, we assume the surfactant to be solvable and, hence, only present in the liquid phase.
Furthermore, we introduce a compressive flux similar to the one in \autoref{eq:transportEqn} for the volume fraction field to counteract numerical diffusion of surfactant into the gas phase. A volumetric sink term in the interface region $j_\mathrm{B}$ is added to account for adsorption to the liquid-gas interface. The transport equation for the volume-averaged bulk concentration $c_\mathrm{B}$ then reads as
\begin{equation}
\frac{\partial c_\mathrm{B}}{\partial \tau} + \nabla\cdot(c_\mathrm{B} \mathbf{u}) + \nabla\cdot(c_\mathrm{B} (1-\alpha) \mathbf{u}_\mathrm{r}) = \nabla\cdot (D \nabla c_\mathrm{B}) - \nabla\cdot\left(\frac{c_\mathrm{B}}{\alpha+\varsigma_\alpha} D \nabla\alpha \right) -j_\mathrm{B}.
\label{eq:bulkFinal}
\end{equation}

Therein, $D$ is the diffusion coefficient within the liquid. The small value $\varsigma_\alpha$ is added to the denominator to avoid numerical issues due to division by zero in the gas phase outside of the interface region\footnote{Within the scope of this work, a value of $10^{-15}$ is used.}.
Thus, the left-hand side of \autoref{eq:bulkFinal} comprises instationary, convective as well as artificial interface compression terms. The right-hand side consists of two diffusive terms arising from conditional volume averaging across both phases and the volumetric sink term representing adsorption of surfactant from the liquid bulk to the liquid-gas interface.
The volume-averaged bulk concentration field thus is zero within the gas phase, where $\alpha = 0$, and it is $\geq 0$ within the liquid phase, where $\alpha=1$. At the liquid-gas interface, it displays a small transition region. The locally averaged bulk concentration within the liquid phase can be calculated according to
\begin{equation}
\overline{c_1}^1 = \frac{c_\mathrm{B}}{\alpha+\varsigma_\alpha}.
\label{eq:c1FromCB}
\end{equation}
Note that $\overline{c_1}^1 = c_\mathrm{B}$ within the liquid bulk, where $\alpha=1$, but the two concentration values differ within the interface region.
In analogy to \autoref{eq:c1FromCB}, $c_1$ will be used throughout this article to denote the concentration in the liquid phase before conditional volume averaging is applied. Thus, $c_1$ denotes the concentration within the liquid phase and $\overline{c_1}^1$ its local volume average in the liquid phase.

\subsubsection{Interfacial Excess Concentration}
The transport of a surfactant present at the liquid-gas interface is generally understood in terms of the surface excess concentration. Specifically, evaluating the surface tension from typical surface equations of state, or adsorption rates from typical sorption models, the area-specific surface excess concentration $\Gamma_\mathrm{\ell g}$ is a most important variable.  As the volume of fluid method lacks an area mesh coinciding with the interface, solving a transport equation for this area-specific surface excess concentration on the interface is not straightforward. To circumvent this issue, a different approach is chosen here. Instead of solving directly for species amount per interfacial area, the area-specific surface excess concentration is transferred to a volume specific surface excess concentration, hereby referred to as interfacial excess concentration $c_\mathrm{I}$. This interfacial excess concentration $c_\mathrm{I}$ is calculated as
\begin{equation}
c_\mathrm{I} = \delta_\mathrm{I} \Gamma_\mathrm{\ell g}.
\label{eq:cIFromGamma}
\end{equation}
This concept has been employed previously in the context of the conservative level-set method by \citet{Lakshmanan2010}. A convection-diffusion equation
\begin{equation}
\frac{\partial c_\mathrm{I}}{\partial \tau}
+ \nabla\cdot(c_\mathrm{I} \mathbf{u})
+ \nabla\cdot\left(c_\mathrm{I} w(\alpha) \mathbf{u}_\mathrm{r} \right)
= %
{\nabla%
\cdot(D_\mathrm{I} \nabla_\Sigma c_\mathrm{I})}
+\nabla\cdot\left(D_{\mathrm{I,S}}\nabla c_\mathrm{I}\right)
+j_\mathrm{I}
\label{eq:interfaceFinal}
\end{equation}
is then used to describe the evolution of this interfacial excess concentration. 
While \citet{Lakshmanan2010} solved an additional reinitialization equation for $c_\mathrm{I}$, here a compressive flux similar to those for $\alpha$ and $c_\mathrm{B}$ is added directly to the transport equation. The heuristically chosen weighting of the compressive flux
\begin{equation}
w(\alpha)=\mathrm{sgn}(\frac{1}{2}-\alpha) (1-4 \alpha (1-\alpha))
\end{equation}
is found to produce good results, as will be demonstrated in \autoref{sec:AdvectionDiffusion}. The diffusive term with $D_\mathrm{I,S}$ ensures sufficient smoothness of $c_\mathrm{I}$ across the interface. The coefficients $D_\mathrm{I,S}=||\mathbf{u}_\mathrm{r}||_2\, D_{\mathrm{I,S},0}$ and $D_{\mathrm{I,S},0}=\num{0.4}\,\Delta x$ are found to produce good results, as will be demonstrated by the verification cases presented in subsections \ref{sec:VerificationAdvectionNormal} and \ref{sec:VerificationConvectionTangential}. Diffusion of surfactant along the interface is taken into account by the first diffusive term on the right-hand side. Adsorption to the interface is modeled through the volumetric source term $j_\mathrm{I}$. Evaluation of $j_\mathrm{I}$ and its relation to $j_\mathrm{B}$ will be presented in detail in \autoref{sec:AdsorptionDesorption} below.

\begin{figure}[t]

\centering
\def\svgwidth{0.9\linewidth}

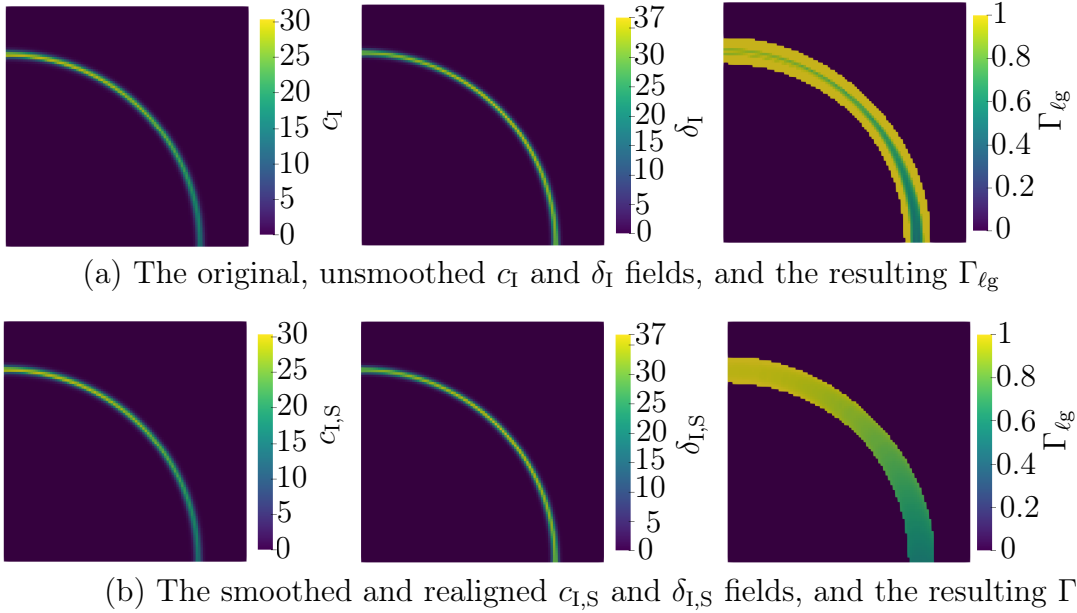

\caption{The necessity of the smoothing and realignment step for the calculation of $\Gamma_\mathrm{\ell g}$ is demonstrated. $c_\mathrm{I,S}$ and $\delta_\mathrm{I,S}$ correspond to the smoothed and realigned $c_\mathrm{I}$ and $\delta_\mathrm{I}$, following the final step during realignment. The units of $c_{\mathrm{I}}$, $c_{\mathrm{I,S}}$ are $\mathrm{mol/m^3 }$, $\delta_{\mathrm{I}}$, $\delta_{\mathrm{I,S}}$ are $\mathrm{1/m}$, and $\Gamma_\mathrm{\ell g}$ are $\mathrm{mol/m^2}$. }
\label{fig:effectReeinitializationGamma}
\end{figure}

\paragraph{Surface excess concentration}
As mentioned before, the area-specific surface excess concentration $\Gamma_{\mathrm{\ell g}}$, hereby referred to as surface excess concentration, is an important variable, governing the variation of local surface tension and also the sorption kinetics in case of soluble surfactants. However, simply rearranging \autoref{eq:cIFromGamma} for $\Gamma_\mathrm{\ell g}$ results in substantial numerical errors. This is demonstrated in \autoref{fig:effectReeinitializationGamma}a, where the numerical variables correspond to the case presented in detail in Section \ref{sec:VerificationAdvectionNormal}. Therein, \autoref{fig:effectReeinitializationGamma}a shows $c_\mathrm{I}$ and $\delta_\mathrm{I}$ together with the corresponding $\Gamma_\mathrm{\ell g}$. It can be seen that $\Gamma_\mathrm{\ell g}$ is not constant in interface normal direction, but is overestimated towards the brim. The reason for this is numerical diffusion acting differently on the Dirac delta-like  $c_\mathrm{I}$ compared to the step like $\alpha$ field on which the local interface density $\delta_\mathrm{I}$ is based. As the interface region is typically resolved with approximately three grid cells \citep{Woerner2012}, independent of grid resolution, an improvement with mesh resolution can not be expected. In order to overcome this issue, we propose a smoothing and realignment procedure. This consists of multiple smoothing and realignment steps, where in each step, a convection-diffusion equation in pseudo-time $\tau^*$ is solved for both interfacial excess concentration and regularized Dirac delta. For the smoothed and realigned interfacial excess concentration $c_\mathrm{I,S}$, the advection-diffusion equation
\begin{equation}
\frac{\partial c_\mathrm{I,S}}{\partial \tau^*} + \nabla\cdot (c_\mathrm{I,S}\, w(\alpha)\, \mathbf{u}_{c,\delta}) = \nabla\cdot (\mathbf{n}_\kappa\otimes\mathbf{n}_\kappa\ D_{c,\delta} \nabla c_\mathrm{I,S})
\end{equation}
is integrated in pseudo time until $\tau^*_\mathrm{S}$, starting from $c_\mathrm{I,S}(\tau^*=0)=c_\mathrm{I}$.
Similarly, the regularized Dirac delta is smoothed and realigned, starting from $\delta_\mathrm{I,S}(\tau^*=0)=\delta_\mathrm{I}$, by integrating
\begin{equation}
\frac{\partial \delta_\mathrm{I,S}}{\partial \tau^*} + \nabla\cdot (\delta_\mathrm{I,S}\, w(\alpha)\, \mathbf{u}_{c,\delta}) = \nabla\cdot (\mathbf{n}_\kappa\otimes\mathbf{n}_\kappa\  D_{c,\delta} \nabla \delta_\mathrm{I,S})
\end{equation}
in pseudo time until $\tau^*_\mathrm{S}$. The area-specific surface excess concentration can then be evaluated from
\begin{equation}
\Gamma_\mathrm{\ell g}=\frac{c_\mathrm{I,S}}{\delta_\mathrm{I,S} + \varsigma_{\nabla\alpha}}.
\label{eq:GammaFromS}
\end{equation}
As $c_\mathrm{I,S}$ and $\delta_\mathrm{I,S}$ are diffused across the interface region, $\Gamma_\mathrm{\ell g}$ evaluated from \autoref{eq:GammaFromS} is available within the entire interface region. \autoref{fig:effectReeinitializationGamma}b shows $\Gamma_\mathrm{\ell g}$ resulting from the smoothed and realigned fields $c_\mathrm{I,S}$ and $\delta_\mathrm{I,S}$, following the smoothing and realignment procedure. The parameters $D_{c,\delta}$ and $\tau^*_\mathrm{S}$ are chosen such that $\mathit{Pe}=\frac{||\mathbf{u}_{c,\delta}||_2\Delta x_{\mathrm{I}}}{D_{c,\delta}}=1$ and $\frac{||\mathbf{u}_{c,\delta}||_2 \tau^*_\mathrm{S}}{\Delta x_{\mathrm{I}}}=4$ with $\mathbf{u}_{c,\delta}=\SI{1}{\m\per\s}\,\mathbf{n}_\mathrm{\kappa}$. The mesh size at the interface is denoted by $\Delta x_{\mathrm{I}}$. Note that $c_\mathrm{I,S}$ and $\delta_\mathrm{I,S}$ are only used for the evaluation of $\Gamma_\mathrm{\ell g}$. Further evolution of $c_\mathrm{I}$ and $\delta_\mathrm{I}$ is described using \autoref{eq:interfaceFinal} and \autoref{eq:transportEqn}.

\subsubsection{Adsorption and Desorption}
The interfacial excess concentration at the liquid gas-interface and the bulk concentration are iteratively coupled through the respective source terms for adsorption and desorption. This coupling is described here. Following this, the model describing adsorption to a solid substrate wall is introduced.

\paragraph{Liquid-gas interface}
The source term for adsorption and desorption at the liquid-gas interface within the proposed method describes the rate of adsorption or desorption in the form of 
\label{sec:AdsorptionDesorption}
\begin{equation}
j_\mathrm{I}\, =\, A\, \delta_\mathrm{I} + B\, c_\mathrm{I}.
\label{eq:interfaceSourceTerms}
\end{equation}
Thus, the sorption source term is described in volume specific form and is non-zero only in the interface region. A similar approach was introduced by \citet{Lakshmanan2010} within the conservative level-set framework. For linear sorption kinetics models, such as Langmuir-Hinshelwood kinetics (see e.g. \citep{Chang1995} for reference), this allows for an implicit treatment of the source terms in \autoref{eq:interfaceFinal}. Non-linear sorption kinetics, such as described by the Langmuir-Freundlich model \citep{Sips1948,Chan2012}, can be brought into the above form by linearization around the current surface excess concentration $\Gamma_\mathrm{\ell g,0}$. An example for Langmuir-Freundlich kinetics is given in \autoref{sec:AppendixFreundlich}. The coefficients then, in general, depend on $\Gamma_\mathrm{\ell g,0}$ as well as on the local sub-surface bulk concentration $c_\mathrm{1,I}$. As both fields are required within the entire interface region, the latter is distributed over the interface. For this purpose, the sub-surface concentration
\begin{equation}
c_{1,\delta_2}=\overline{c_\mathrm{1}}^1 \delta_2=\frac{c_\mathrm{B}}{\alpha+\varsigma_\alpha}\delta_2
\label{eq:c1Delta2}
\end{equation}
is evaluated and distributed over the interface region. This distribution is achieved by solving the diffusion equations 
\begin{equation}
\frac{\partial c_{1,\delta_2,S}}{\partial \tau^*}=D_\mathrm{1,I,S} \nabla^2 c_{1,\delta_2,S}
\end{equation}
and
\begin{equation}
\frac{\partial \delta_\mathrm{2,S}}{\partial \tau^*}=D_\mathrm{1,I,S} \nabla^2 \delta_\mathrm{2,S}
\end{equation}
in pseudo time and finally evaluating the distributed sub-surface bulk concentration at $\tau^*_\mathrm{max}$ as
\begin{equation}
c_\mathrm{1,I}=\frac{c_{1,\delta_2,S}}{\delta_\mathrm{2,S}+\varsigma_{\nabla\alpha}}.
\label{eq:cBulkAtInterface}
\end{equation}
Similar to the realignment for the interfacial excess concentration, $D_\mathrm{1,I,S}$ and $\tau^*_\mathrm{max}$ are chosen such that $\frac{D_\mathrm{1,I,S}\tau^*_\mathrm{max}}{\Delta x^2}=4$.

\begin{figure}
\centering
\def\svgwidth{0.9\linewidth}

\import{plotData/VerificationValidation/jSourceDistribution/}{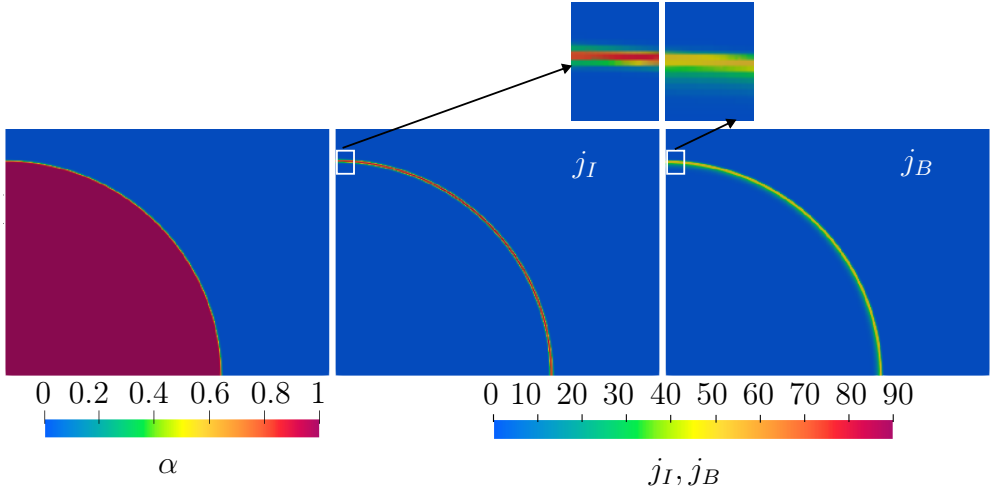}
\caption{Liquid phase volume fraction $\alpha$ and the source term $j_{\mathrm{I}}$ calculated from the sorption kinetics are shown. This $j_{\mathrm{I}}$ is integrated in pseudo time to obtain the distributed bulk source term $j_{\mathrm{B}}$. Insets show a zoomed-in image of a part of the liquid-gas interface, highlighting the shift of source terms towards the liquid phase. The units for $j_{\mathrm{I}}$ and $j_{\mathrm{B}}$ are $\mathrm{mol/m^{3}s}$.}
\label{fig:jBdistribution}
\end{figure}

The coupling of the transport equations for $c_\mathrm{B}$ and $c_\mathrm{I}$ requires that $j_\mathrm{I}$ must be equal to $j_\mathrm{B}$ and should enter the transport equation for $c_\mathrm{B}$. However, applying the sorption source term $j_\mathrm{I}$ directly to the transport equation for bulk concentration within the narrow interface region is numerically problematic, as it may result in negative values of $c_\mathrm{B}$ towards the gas phase of the interface. For this reason, the position at which the source term $j_\mathrm{B}$ is applied is shifted slightly into the liquid phase. To obtain the shifted bulk source term $j_\mathrm{B}$, once more an advection-diffusion equation is integrated in pseudo time until $\tau^*_\mathrm{max}$, where $j_\mathrm{B}(\tau^*=0)=j_\mathrm{I}$. This transport equation
\begin{equation}
\frac{\partial j_\mathrm{B}}{\partial \tau^*}
+ \nabla\cdot(j_\mathrm{B} (1-\alpha) \mathbf{u}_\mathrm{r})
= \nabla\cdot (D_{j,\mathrm{B}} \nabla j_\mathrm{B})
- \nabla\cdot\left(\frac{j_\mathrm{B}}{\alpha+\varsigma_\alpha} D_{j,\mathrm{B}} \nabla\alpha \right)
\label{eq:bulkSourceDistribution}
\end{equation}
includes a compressive and diffusive terms similar to the ones in \autoref{eq:bulkFinal} in order to allow diffusion only within the liquid phase.
The parameters are again chosen such that $\frac{D_{j,\mathrm{B}}\tau^*_\mathrm{max}}{\Delta x^2}=4$. Note that, similar to $j_{\mathrm{I}}$, the term $j_{\mathrm{B}}$ is also a volume specific source term, as visible from the transport equations for $c_{\mathrm{B}}$ and $c_{\mathrm{I}}$, in \autoref{eq:bulkFinal} and \autoref{eq:interfaceFinal}, respectively.  For an exemplary scenario of adsorption occurring at a liquid-gas interface, \autoref{fig:jBdistribution} shows the $\alpha$ field, the source term $j_{\mathrm{I}}$ and the distributed bulk source term $j_{\mathrm{B}}$ at a particular time instant. This numerical case is further presented in detail in \autoref{fig:verificationAdsorption2DPesciSlow} of Section \ref{sec:AdsorptionatLGinterface}. As can be expected, $j_{\mathrm{I}}$ is concentrated in the region of liquid-gas interface and $j_{\mathrm{B}}$ is shifted towards the liquid phase.

\paragraph{Solid-fluid interface}
\label{sec:MethodAdsorptionatSLinterface}
Sorption processes occurring at a solid-fluid interface can be described by an appropriate boundary condition for $c_\mathrm{B}$.  The term solid-fluid interface highlights the possibility that the phase adjacent to the solid may be either liquid or gas.
Furthermore, two (or more) fluid phases may be in contact with the solid wall, resulting in one (or more) three-phase contact lines.
In our case, where a liquid-gas system is considered, this results in three distinct interfaces as solid-liquid, liquid-gas and solid-gas interface. Therefore, the solid surface consists of a solid-liquid and a solid-gas interface, which is the typical situation at a wetted wall. Throughout this work, the term "wall" indicates the surface of the solid in contact with fluid.

At a wall, the net diffusive flux of surfactant must equal the area-specific adsorption rate $j_\mathrm{s}$. Taking both diffusive terms of the volume averaged conservation equation for $c_\mathrm{B}$ ( \autoref{eq:bulkFinal}) into account, the boundary condition follows as
\begin{equation}
\mathbf{n}_\mathrm{w}\cdot D\nabla c_\mathrm{B} - \frac{c_\mathrm{B}}{\alpha+\varsigma_\alpha}\, \mathbf{n}_\mathrm{w}\cdot D\nabla\alpha = j_\mathrm{s}.
\label{eq:boundarycondition}
\end{equation}
Therein, $\mathbf{n}_\mathrm{w}$ is the wall normal vector. It can be seen that, for a non-zero $c_{\mathrm{B}}$, the second term on the left hand side is non-zero only if there exists a gradient in $\alpha$ perpendicular to the wall. This precisely occurs at the region where the liquid-gas interface meets the solid wall, i.e. at a three-phase contact line. Accordingly, away from the three-phase contact line, the boundary condition is reduced to an inhomogeneous Neumann condition. The area-specific adsorption rate at the solid-fluid interface is evaluated as 
\begin{equation}
j_\mathrm{s} = \dot{\Gamma}_\mathrm{s} = \alpha j_\mathrm{s\ell},
\label{eq:balanceGammas}
\end{equation}
where $\Gamma_\mathrm{s}$ is the surface excess concentration at the solid-fluid interface. Further, $j_\mathrm{s\ell}$ is the area-specific adsorption rate at the solid-liquid interface, and may be described as a function of the current local bulk concentration $\overline{c_\mathrm{1}}^1$ within the liquid, and the surface excess concentration at the solid-liquid interface $\Gamma_\mathrm{s\ell}$, using appropriate kinetic models such as Henry, Langmuir-Hinshelwood, or Langmuir-Freundlich kinetics. Langmuir-Freundlich kinetics are presented in \autoref{sec:AppendixFreundlich} as a frequently employed example. Moreover, it is to be noted that the terms $j_{\mathrm{s}}$ and $j_{\mathrm{s\ell}}$ in \autoref{eq:balanceGammas} denote area-specific adsorption rates, whereas, for the liquid-gas interface, the terms $j_{\mathrm{I}}$, $j_{\mathrm{B}}$ in \autoref{eq:bulkFinal} and \autoref{eq:interfaceFinal} indicate volume specific adsorption rates.

For cases in which the wetted area is fixed, i.e.\ for an immobile contact line, $\Gamma_\mathrm{s\ell}$ is only impacted by the rate of adsorption at the solid-liquid interface. However, for a case with increasing wetted area, for instance at an advancing contact line, along with the rate of adsorption, $\dot{\Gamma}_\mathrm{s\ell}$ depends also on any surfactant that was present in the differential area over which the liquid advances. Due to the advancing contact line, surfactant remaining at a fixed adsorption site on the substrate may change from being adsorbed to the solid-gas interface to the solid-liquid interface. Thus, the moving contact line may change the average surface excess concentrations at solid-gas and solid-liquid interfaces within the control area.
Therefore, aside from keeping track of the surface excess concentration of surfactant at the solid-fluid interface $\Gamma_\mathrm{s}$, there is a need to keep track of the surface excess concentration of surfactant  both at the solid-liquid interface ($\Gamma_\mathrm{s\ell}$), and the solid-gas interface ($\Gamma_\mathrm{sg}$), which are then related to each other as
\begin{equation}
\Gamma_\mathrm{s} = \alpha \Gamma_\mathrm{s\ell} + (1-\alpha) \Gamma_\mathrm{sg}.
 \label{eq:GammaAverage}
\end{equation}

Thus, similar to $\Gamma_{\mathrm{\ell g}}$, the surface excess concentrations at the solid wall $\Gamma_{\mathrm{s}}$, $\Gamma_{\mathrm{s\ell}}$, and $\Gamma_{\mathrm{sg}}$ are area-specific quantities. 

\begin{figure}

\centering
\def\svgwidth{0.9\linewidth}

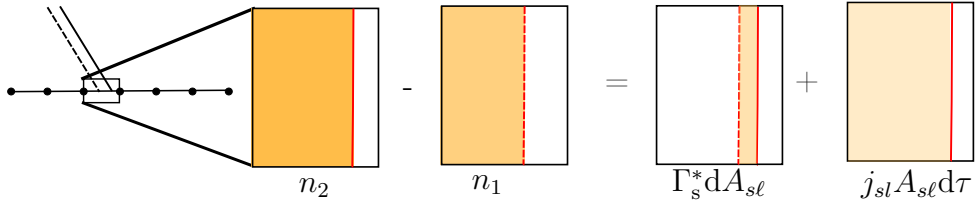
\caption{Schematic showing the balance of amount of surfactant adsorbed within a controlled area when the contact line advances, resulting in an increased wetted area. The initial contact line at time $\tau$ is shown as a dotted line and the contact line at time $\tau+\mathrm{d}\tau$ is shown as a continuous line. $n_{2}$ and $n_{1}$ denote the amount of species adsorbed at the solid at time $\tau$ and $\tau+\mathrm{d}\tau$, respectively.}
\label{fig:speciesbalance}
\end{figure}

At any instant, with $A_{\mathrm{s\ell}}$ being the wetted part in a differentially small control area, (see \autoref{fig:speciesbalance}), let $n$ be the amount of surfactant adsorbed within this control area, i.e. $n = \Gamma_\mathrm{s\ell}\,A_{\mathrm{s\ell}}$. Accordingly, mass conservation dictates that the change in the amount of surfactant adsorbed within a control area can be written as 

\begin{equation}
\mathrm{d}n = \Gamma_\mathrm{s}^*\, \mathrm{d}A_\mathrm{s\ell} + j_\mathrm{s\ell} \,A_\mathrm{s\ell}\, \mathrm{d}\tau,
\label{Eq:Balance}
\end{equation}
where $\Gamma_\mathrm{s}^*$ is the surface excess concentration in the area by which the control area is increased. Therefore, the first term on the right-hand side is the increase in adsorbed amount due to the increase of the control area (i.e. wetted area) towards a region already covered with surfactant. The second term accounts for the amount of surfactant adsorbed from the liquid bulk phase. \autoref{fig:speciesbalance} shows a schematic for this balance of surfactant amount for an advancing contact line. This balance furthermore assumes that there is no surfactant transport along the substrate and no other sources or sinks, e.g. due to chemical reactions.

With

$\mathrm{d}n = A_\mathrm{s\ell}\,\mathrm{d}\Gamma_\mathrm{s\ell} + \Gamma_\mathrm{s\ell}\,\mathrm{d}A_\mathrm{s\ell}$, it follows from \autoref{Eq:Balance} that
\begin{equation}
\mathrm{d}\Gamma_\mathrm{s\ell}=\frac{\Gamma_\mathrm{s}^* - \Gamma_\mathrm{s\ell}}{A_\mathrm{s\ell}}\,\mathrm{d}A_\mathrm{s\ell} + j_\mathrm{s\ell}\,\mathrm{d}\tau
\end{equation}
and thus
\begin{equation}
\dot{\Gamma}_\mathrm{s\ell}=\frac{\Gamma_\mathrm{s}^* - \Gamma_\mathrm{s\ell}}{A_\mathrm{s\ell}}\,\dot{A}_\mathrm{s\ell} + j_\mathrm{s\ell}.
\end{equation}
Assuming that within this control area, the fraction of substrate surface area covered by liquid is equal to the liquid volume fraction at the solid wall 
\begin{equation}
\frac{A_\mathrm{s\ell}}{A_\mathrm{s\ell}+A_\mathrm{sg}}=\alpha_\mathrm{w},
\end{equation}
this balance can also be expressed as 
\begin{equation}
\dot{\Gamma}_\mathrm{s\ell}=\frac{\Gamma_\mathrm{s}^* - \Gamma_\mathrm{s\ell}}{\alpha_\mathrm{w}}\,\dot{\alpha}_\mathrm{w} + j_\mathrm{s\ell}.
\end{equation}
Furthermore, depending on whether the contact line is advancing or receding, $\Gamma_\mathrm{s}^*$ corresponds either to $\Gamma_\mathrm{sg}$ or $\Gamma_\mathrm{s\ell}$. Again introducing the regularization parameter $\varsigma_\alpha$ to avoid division by zero at the solid-gas interface, we can write the change in local surface excess concentration at the solid-liquid interface as
\begin{equation}
\dot{\Gamma}_\mathrm{s\ell}=
\begin{cases}
 j_\mathrm{s\ell} + \frac{\Gamma_\mathrm{sg} - \Gamma_\mathrm{s\ell}}{\alpha_\mathrm{s} + \varsigma_\alpha}\,\dot{\alpha}_\mathrm{w} & \text{for $\dot{\alpha}_\mathrm{w} > 0$} \\

 j_\mathrm{s\ell} & \text{for $\dot{\alpha}_\mathrm{w} \leq 0$}.
\end{cases}
\label{eq:balanceGammasld}
\end{equation}
From \autoref{eq:balanceGammasld}, it can be observed that, when the wetted area is constant or decreases (for a receding contact line), $\dot{\Gamma}_{\mathrm{s\ell}}$ is only impacted by $j_{\mathrm{s\ell}}$. Thereby, \autoref{eq:balanceGammasld} accounts for the influence of contact line motion on the surface excess concentration fields. Thus, the evolution of $\Gamma_\mathrm{s}$, $\Gamma_\mathrm{s\ell}$ and $\Gamma_\mathrm{sg}$ is described through \autoref{eq:balanceGammas}, \autoref{eq:GammaAverage} and \autoref{eq:balanceGammasld}.

\subsection{Numerical Methodology}
For the numerical solution of the above described model, the governing equations are discretized using the schemes listed in \autoref{sec:schemes}. For the volume fraction field as well as for the concentration fields, it is important to maintain boundedness. For this purpose, the self-filtered central differencing (SFCD) and van Leer \citep{vanLeer1974} schemes as implemented in OpenFOAM \citep{OpenFOAM2016,OpenFOAM2018} are used. Special care has to be taken also with respect to the terms arising from conditional volume averaging of the diffusive terms \citep{Deising2016}. To ensure consistency between the respective Laplacian and divergence terms, the divergence terms are discretized using central differencing, while the cell face-normal gradient used in the discretization of the Laplacian uses central differencing across the cell face as well.\footnote{The non-orthogonal correction of the limited scheme vanishes for the orthogonal meshes considered in this article.} Similarly, the cell face-normal gradient of the volume fraction field is discretized and linear interpolation of the volume fraction field to the cell faces is employed for the divergence term arising from diffusion of the bulk concentration.

\begin{figure}[htbp]
\tikzset{%
  >={Latex[width=2mm,length=2mm]},
            base/.style = {rectangle, draw=black,rounded corners,
                           minimum width=6cm, minimum height=1cm,
                           text centered, font=\sffamily},
  activityStarts/.style = {base,minimum width=1cm,ellipse, fill=green!10},
       startstop/.style = {base, fill=red!30},
    activityRuns/.style = {base, fill=green!30},
 		decision/.style = {diamond,fill=orange!20, draw=black,aspect=4,minimum width=4cm, text width=5em},
         process/.style = {base, minimum width=2.5cm, fill=yellow!20},
         groupedProcess/.style = {base, minimum width=2.5cm, fill=gray!20},
}
\centering
\includegraphics{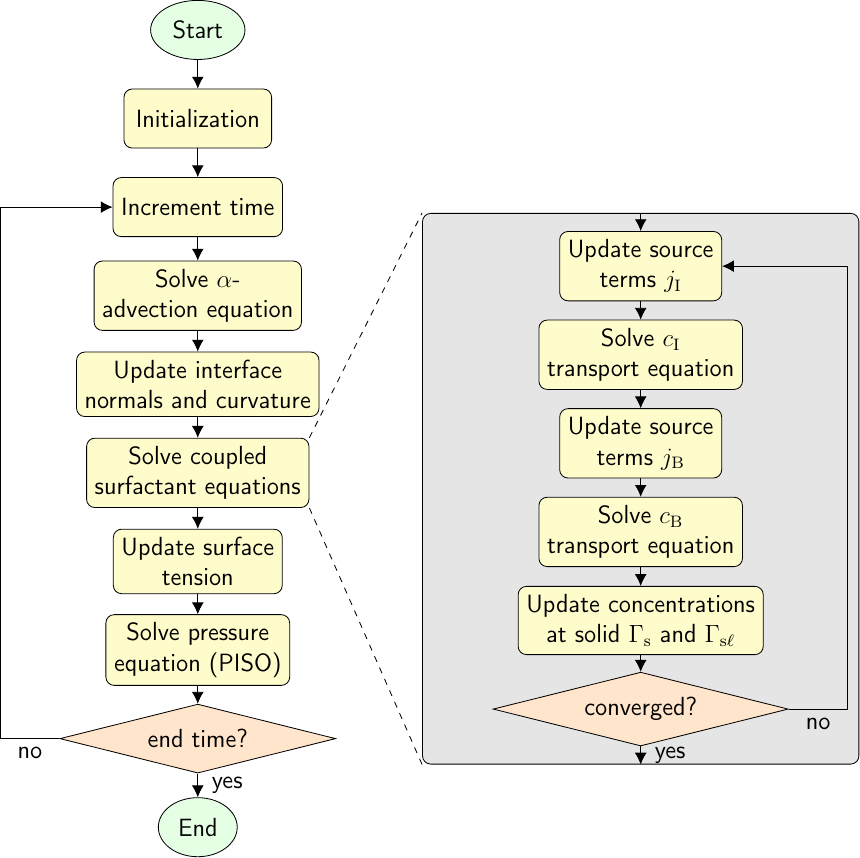}
\caption{Solution procedure.}
\label{fig:solutionProcedure}
\end{figure}

Our method describing soluble surfactants is implemented into the \emph{interFoam} solver of the open source CFD library OpenFOAM \citep{Greenshields2016,Greenshields2018} extended by several in-house developments regarding heat transfer, capillary forces, as well as adaptive mesh refinement and load-balancing \citep{Kunkelmann2011,Batzdorf2015,Rettenmaier2019Diss,Franz2021Diss}. However, in the following, focus will be only on the aspects relevant to the present work.
\autoref{fig:solutionProcedure} shows the solution procedure. After initialization of mesh and fields, first the advection equation for the volume fraction field is solved. Based on the new volume fraction field, the interface normals and interface curvature are updated. Following this, surfactant transport is handled. Bulk and interfacial excess concentrations are therefore iteratively coupled.
For cases with adsorption to the solid substrate, the boundary condition for the bulk concentration field $c_\mathrm{B}$ at the solid-fluid interface, \autoref{eq:boundarycondition}, is implemented as a mixed boundary condition within OpenFOAM, as presented in \autoref{sec:BoundaryCondition}.
Based on the new surface excess concentrations, surface tension and surface tension forces are updated, which enter the momentum balance. Finally, the pressure field is updated using the PISO algorithm \citep{Issa1986} at the end of each time step.

\section{Verification}
\label{sec:Verification}
In order to verify the above presented comprehensive numerical methods, we consider several test cases. First, for an insoluble surfactant, the transport of surfactant with the liquid-gas interface and along the liquid-gas interface will be investigated in \autoref{sec:AdvectionDiffusion}. Second, for a soluble surfactant, the coupling between bulk and interface concentrations is evaluated in \autoref{sec:VerificationAdsorptionDesorption} in order to verify the methodology describing ad- and desorption processes at the liquid-gas and solid-fluid interfaces. Finally, the back-coupling of surfactant transport to the momentum balance via Marangoni-stresses is evaluated in \autoref{sec:MarangoniFlow}.
\subsection{Advection and Diffusion}
\label{sec:VerificationConvection}
The verification with respect to transport of surfactant with and along the interface is further devided into advective transport in interface normal direction (section \ref{sec:VerificationAdvectionNormal}), advective and diffusive transport in tangential direction (section \ref{sec:VerificationConvectionTangential})).
\label{sec:AdvectionDiffusion}
\subsubsection{Advection in interface normal direction}
\label{sec:VerificationAdvectionNormal}
\begin{figure}
\begin{subfigure}{0.99\textwidth}
	\includegraphics{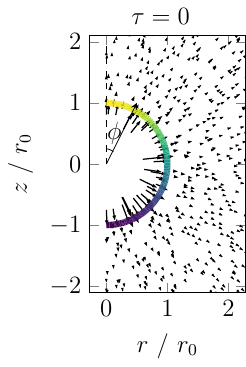}
		\includegraphics{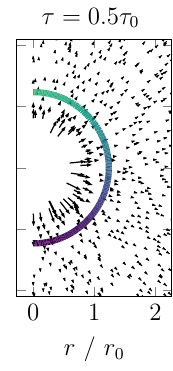}
		\includegraphics{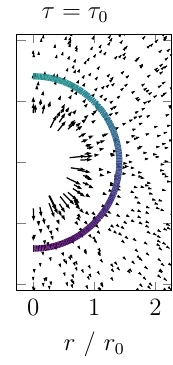}
			\includegraphics{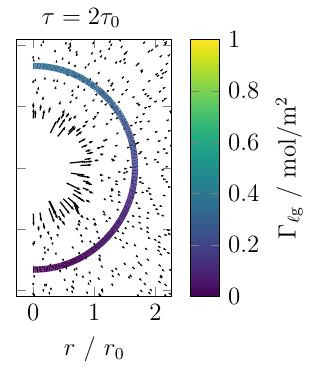}
	\caption{ The velocity field and the surface excess concentration at the liquid-gas interface, at different time instants are shown. The time $\tau_0$ corresponds to the time when the surface area of the drop is twice that that at the initial time, $\tau=0$.}
	\label{fig:RadialAdvectionFields}
	\end{subfigure}
	
	\begin{subfigure}[t]{.45\textwidth}
	\centering
	\includegraphics{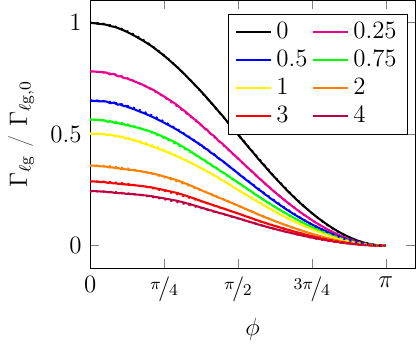}
	\caption{The surface excess concentration from numerical solution (solid lines) is compared to that from the analytical reference solution (dashed lines), at different instants of $\tau/\tau_0$. }
	\label{fig:ComparisonatDifferentTimesRadialAdvection}
\end{subfigure}
\begin{subfigure}[t]{.45\textwidth}
	\centering
	\includegraphics{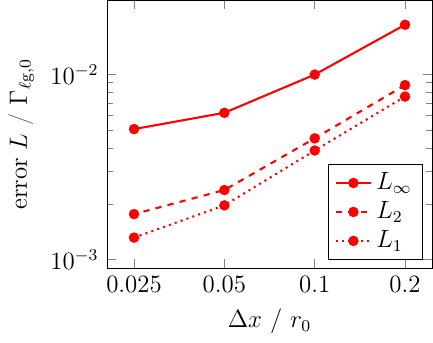}
	\caption{ The influence of mesh resolution on error norms is shown at a time instant of $\tau = 4\tau_0$.}
	\label{fig:ErrorRadialAdvectionError}
\end{subfigure}%

\caption{Results from the verification case of an expanding spherical drop, advecting the surfactant present at the liquid-gas interface in normal direction.}
\end{figure}
For the verification of surfactant transport in interface normal direction, an expanding, spherical drop is considered. Aside from the chosen parameters for diameter and expansion rate, this test case is identical to the one introduced by \citet{James2004} to verify their method for the simulation of transport of insoluble surfactant based on geometric VOF. Using cylindrical coordinates, the point symmetric velocity field driving the expansion is given by  
\begin{equation}
\textbf{u} = \frac{r\,\mathbf{e}_r + z\, \mathbf{e}_z}{(r^2+z^2)^\frac{3}{2}}\ \dot{V}_0,\qquad r,z >0.
\end{equation}
Therein, $r$ and $z$ represent the radial and axial coordinates, respectively.
For the purpose of this verification case, the velocities are assumed to be time-independent.
The drop radius, therefore, increases according to
\begin{equation}
r(\tau) = \sqrt[3]{3\dot{V}_0\tau + r_0^3},
\end{equation}
where $r(\tau=0)=r_0=\SI{1}{\m}$. The volume source at the drop's center is given by $\dot{V}_0=\SI{1}{\cubic m\per s}$. With an initial drop radius of $r_0=\SI{1}{\m}$ and $\Gamma_0 = \SI{1}{mol\per\square\m} $, the initial surface excess concentration is set to
\begin{equation}
\Gamma_{\mathrm{\ell g}}(\phi, \tau=0) = \frac{\cos(\phi)+1}{2}\, \Gamma_{\mathrm{\ell g,0}},
\end{equation}
where $\phi$ is the angle enclosed between a point on the drops surface and the axial coordinate. The initial surface excess concentration along with the velocity field is shown in \autoref{fig:RadialAdvectionFields}, at $\tau=0$.

The numerical mesh for this test case makes use of the axial symmetry of the problem. The spatial discretization is varied between mesh sizes of $\num{0.025}\,r_0$ and $\num{0.2}\,r_0$. To avoid the point source at the drop's center, a small region is excluded from the computational domain, and inlet velocities are specified according to the above velocity field.

\autoref{fig:RadialAdvectionFields} shows the surface excess concentration at different time instants, as predicted by the numerical model. While the drop expands, the local concentration on the interface decreases. This is expected, as the surface area increases, while the surfactant amount at the interface remains constant. At the same time, the concentration gradient along the interface is maintained.

For a quantitative evaluation, it is useful to compare the numerical results with the analytical solution.  Without transport in tangential direction, the surface excess concentration evolves according to
\begin{equation}
\Gamma_{\mathrm{\ell g}}(\phi, r(\tau)) = \frac{\cos(\phi)+1}{2}\, \Gamma_{\mathrm{\ell g,0}} \ \frac{r_0^2}{r(\tau)^2}.
\end{equation}

\autoref{fig:ComparisonatDifferentTimesRadialAdvection} shows the surface excess concentration along the angular coordinate $\phi$ at different time instants. The numerical results, represented by solid lines, are obtained on the finest considered grid with $\Delta x=\num{0.025}\,r_0$. The analytical reference solution is shown by dotted lines. The comparison shows that the evolution of surface excess concentration expected from the analytical solution is excellently captured by the numerical results. The influence of mesh resolution on the accuracy of the results is shown in \autoref{fig:ErrorRadialAdvectionError} in terms of $L_1$, $L_2$ and $L_\infty$ error norms, which are calculated from the surface excess concentration at the $\alpha=\num{0.5}$ iso-surface as
\begin{equation}
L_q = 
\left(\sum_{i=1}^N \frac{(|\Gamma_{\mathrm{\ell g,}i}-\Gamma_\mathrm{\ell g}(\phi,r(\tau))|)^q}{N}\right)^{\frac{1}{q}}\qquad \text{for}\ q=1,2,
\end{equation}
and
\begin{equation}
L_\infty=\max_{i=1,\dots,N}\,|\Gamma_{\mathrm{\ell g,}i}-\Gamma_\mathrm{\ell g}(\phi,r(\tau))|.
\end{equation}
Therein, $N$ is the number of sample points and $\Gamma_{\mathrm{\ell g,}i}$ is the surface excess concentration at sample point $i$ from the numerical solution.
In all error norms, the error decreases with mesh resolution. The mesh convergence is roughly of first order. This is expected, as the total variation diminishing (TVD) discretization scheme employed here for surfactant advection is known to reduce to first order accuracy near extrema \citep{Jasak1999}. For $\Delta x / r_0 \leq \num{0.05}$ the relative error is below $1\%$ in all considered error norms. Thus, the surface excess concentration is captured well at reasonable resolution of the liquid-gas interface.

\begin{figure}
\centering
\begin{subfigure}{0.99\textwidth}
\includegraphics{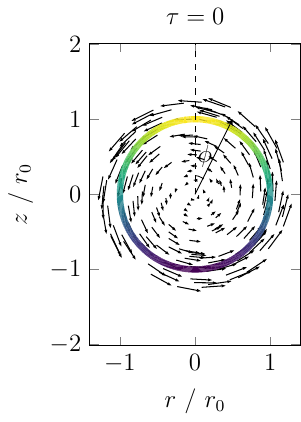}
\includegraphics{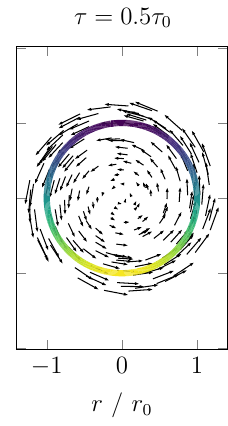}
\includegraphics{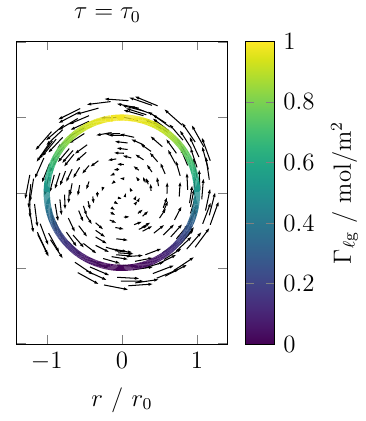}
	\caption{Velocity field and the surface excess concentration from the numerical simulations.The time $\tau_0$ corresponds to the time after one full rotation around the axis of symmetry.}
	\label{fig:TangentialAdvectionFields}
\end{subfigure}

\begin{subfigure}[t]{.45\textwidth}
	\centering
	\includegraphics{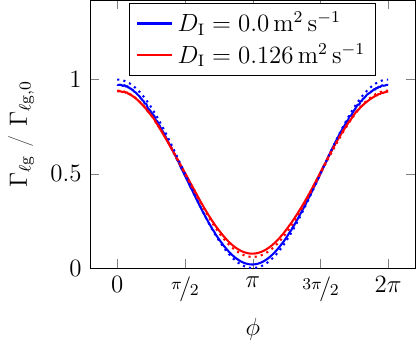}
	\caption{The surface excess concentration from the numerical solution (solid lines) is compared to that from the analytical reference solution (dashed lines), at a time instant of $\tau = \tau_0$,  }
	\label{fig:ErrorTangentialAdvectionGamma}
\end{subfigure}
\begin{subfigure}[t]{.45\textwidth}
	\centering
	\includegraphics{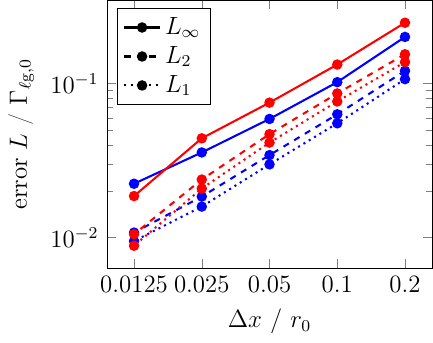}
	
	\caption{At $\tau = \tau_0$, the influence of mesh resolution on the error norms is shown.}
	\label{fig:ErrorTangentialAdvectionError}
\end{subfigure}%

\caption{ Results from the verification case of a rotating disc, advecting the surfactant present at the liquid-gas interface in interface tangential direction. }
\label{fig:ErrorTangentialAdvection}
\end{figure}

\subsubsection{Advection in Interface Tangential Direction}
\label{sec:VerificationConvectionTangential}
Here, the verification of the proposed numerical method for surfactant transport along the interface is presented. For this purpose, the two-dimensional test case of a rotating circular disk with radius $r_0=\SI{1}{\m}$ and center at $x=z=0$ is considered. The disk is rotated by the constant, axisymmetric velocity field 
\begin{equation}
\textbf{v} = (x\, \mathbf{e}_z - z\, \mathbf{e}_x)\ \dot{\phi},
\end{equation}
which is again assumed fixed.
Here, a constant angular velocity of $\dot{\phi}=2\pi\,\SI{}{\per\s}$ is assumed. The initial distribution of the species on the interface is inhomogeneous and given as 
\begin{equation}
\Gamma_{\mathrm{\ell g}}(\phi, \tau=0) = \frac{\cos(\phi)+1}{2}\, \Gamma_{\mathrm{\ell g,0}}.
\end{equation}
The initial velocity field and the surface excess concentration fields are shown in \autoref{fig:TangentialAdvectionFields}, at $\tau=0$. Regarding transport along the interface, two different cases are studied. The first case considers pure advection, thus the interfacial diffusion coefficient vanishes, i.e. $D_\mathrm{I}=0$. In addition, a case with both advection and diffusion along the interface is considered. For this second case an interfacial diffusion coefficient of $D_\mathrm{I}=\SI{0.12566}{\square\m\per\s}$, corresponding to a Peclet number of $\mathit{Pe}_D=100$, is assumed.
Similar cases for the verification with respect to diffusion along the interface have previously been presented by \citet{Muradoglu2008} for the front-tracking method and by \citet{CLERETDELANGAVANT2017271} for the level-set method. However, those cases studied diffusion exclusively without advection in tangential direction. %
For the present case, the domain is discretized using a 2D mesh, where the mesh width is varied between $\num{0.0125}\,r_0$ and $\num{0.2}\,r_0$.

\autoref{fig:TangentialAdvectionFields} shows the evolution of the surface excess concentration with time for the case with pure advection ($D_\mathrm{I}=0$), for the finest grid resolution ($\Delta x=\num{0.0125}\,r_0$). It can be seen that the initial surfactant concentration is rotated around the disc's center with time, while the concentration gradient along the interface is maintained. After one full rotation, the concentration field shows excellent agreement with the initial configuration.

Based on the initial concentration and the prescribed velocity field, the concentration field evolves according to 
\begin{equation}
\Gamma_{\mathrm{\ell g}}(\phi, \tau) = \frac{e^{-D_\mathrm{I}\tau/r_{0}^2}\cos(\phi - \dot{\phi}\tau)+1}{2}\, \Gamma_{\mathrm{\ell g},0}.
\end{equation}
\autoref{fig:ErrorTangentialAdvectionGamma} shows the numerical solution after one full rotation obtained on the finest grid in comparison with the analytical reference solution. Both considered scenarios, i.e. with and without diffusion, show good agreement between the numerical solution and the analytical reference solution. For both cases, it can be observed that the numerical solution slightly overestimates the minimum value, while the maximum value is slightly underestimated. This effect can be attributed to numerical diffusion along the interface. Furthermore, small  spatial oscillations in the numerical solution can be observed around the maximum value. This effect, called 'waviness' \citep{Leonard1991}, was previously observed when central differencing for spatial discretization and explicit discretization for temporal terms \citep{Jasak1999} were used. The influence of different advection schemes on the (1D) advection of $\sin^2$ and semi-ellipse profiles has been studied by \citet{Jasak1999}. For these cases no such oscillations are reported for self-filtered central differencing (SFCD) and van Leer schemes. Since the $\sin^2$ and semi-ellipse profiles are similar to the $\delta$-like $c_\mathrm{I}$ field, it thus seems that advection in interface tangential direction poses a special challenge with respect to advection schemes. Nevertheless, the overall errors remain small even near the maximum value. \autoref{fig:ErrorTangentialAdvectionError} shows the relative error in surface excess concentration for different mesh resolutions. It can be observed that the error decreases with increasing mesh refinement and the mesh convergence is roughly of first order. Again, this is expected, as the employed schemes reduce to first order accuracy near extrema \citep{Jasak1999}.

\subsection{Adsorption}
\label{sec:VerificationAdsorptionDesorption}
Verification of the sorption model, which couples bulk and interfacial excess concentration fields, is presented here in two sections. First, the method is evaluated with respect to adsorption to the liquid-gas interface in section \ref{sec:AdsorptionatLGinterface}. Second, verification of adsorption to the solid-fluid interface, i.e. adsorption to a wall will be discussed in section \ref{sec:AdsorptionatSLinterface}.

\subsubsection{Adsorption to a liquid-gas interface}
\label{sec:AdsorptionatLGinterface}
The verification of the model with respect to adsorption to the liquid-gas interface is based on two test-cases introduced by \citet{PesciDiss2019}, following previous similar studies by \citet{Muradoglu2008}. The two cases focus on adsorption processes (partially) limited by adsorption kinetics (mixed kinetics case) on the one hand, and purely diffusion limited adsorption on the other hand. Both cases describe adsorption of surfactant from the bulk of a spherical drop to its surface. The drop is assumed to be at rest.

\paragraph{Adsorption at the surface of a spherical drop: Mixed kinetics}
Similar to the previous cases, a drop with radius $r_0=\SI{1}{\m}$ is considered. The initial bulk concentration is constant throughout the drop and has a value of $c_1(\tau{=}0,r)=c_{1,0}=\SI{1}{\mol\per\cubic\m}$. Initially, there is no excess surfactant at the interface, thus the initial surface excess concentration $\Gamma_\mathrm{\ell g}(\tau{=}0)=\Gamma_{\mathrm{\ell g},0}=0$. For this test case the simple linear adsorption kinetics model
\begin{equation}
\frac{\partial \Gamma_\mathrm{\ell g}}{\partial \tau} = k_\mathrm{ad,H} c_1|_{r=r_0}
\label{eq:linearKinetics}
\end{equation}
is assumed, facilitating comparison to an analytical reference. An adsorption rate coefficient of $k_\mathrm{ad,H}=\SI{1}{\m\per\s}$ and a bulk diffusion coefficient of $D=\SI{1}{\square\m\per\s}$ are considered. The parameters were chosen according to \citep{PesciDiss2019}. With this parameter choice, curvature effects are of relevance, as the radius of curvature $r_0$ is in the same order of magnitude as the characteristic length scale of adsorption, the so-called adsorption-depth \citep{Ferri2000}, $h_\mathrm{ad}=\Gamma_\mathrm{\ell g}/c_{1}$. %
The numerical solution for this problem is obtained on a 2D axisymmetric domain of size $\num{1.5}\, r_0\times \num{1.5}\, r_0$, discretized with $300\times 300$ cells.

\begin{figure}[h!]
\centering
\begin{subfigure}[t]{\textwidth}
\def\svgwidth{\linewidth}
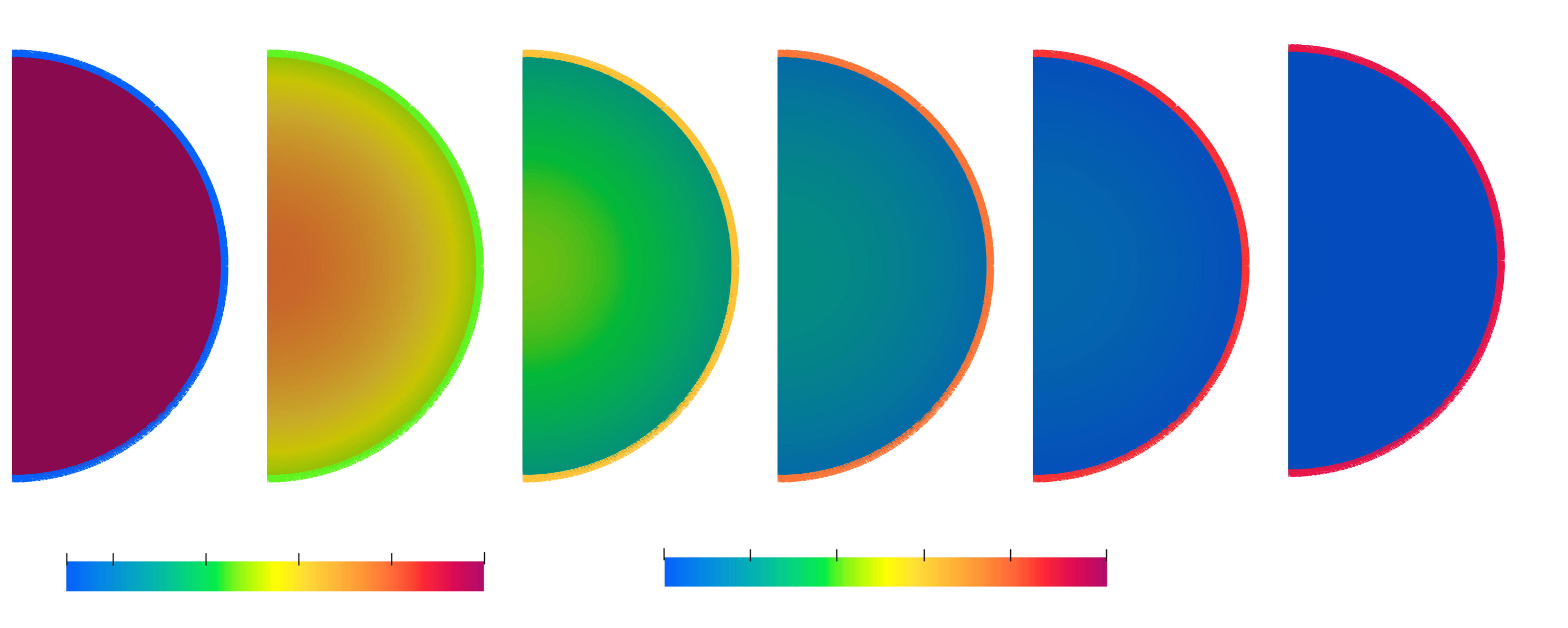
\caption{Bulk concentration and surface excess concentration from the numerical simulation.}
	\label{fig:verificationAdsorption2DPesciSlowContour}
\end{subfigure}
\begin{subfigure}[t]{.45\textwidth}
	\centering
	\includegraphics{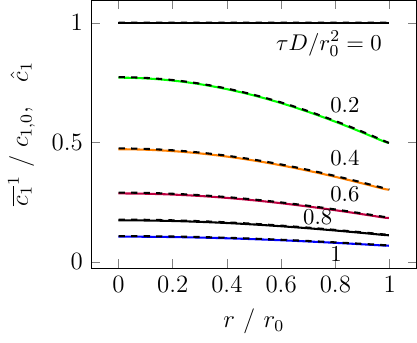}
	\caption{Bulk concentration from numerical solution (colored solid lines) in comparison to the analytical reference solution (black dashed lines) \citep{Antritter2022}.}
		\label{fig:verificationAdsorption2DPesciSlowReference}
\end{subfigure}
\begin{subfigure}[t]{.45\textwidth}
	\centering
\includegraphics{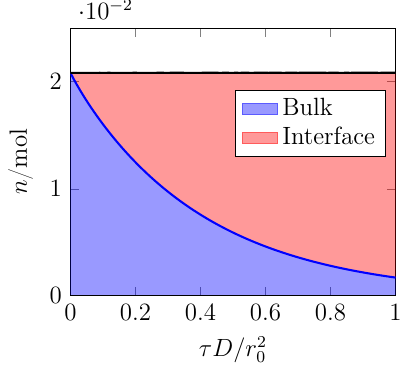}
\caption{Species amount within the liquid bulk and along the interface from numerical solution.}
	\label{fig:verificationAdsorption2DPesciSlowAmount}
\end{subfigure}

		\caption{Results from the verification case for mixed-kinetics adsorption from the bulk of a spherical drop to its interface.  }

	\label{fig:verificationAdsorption2DPesciSlow}
\end{figure}

\autoref{fig:verificationAdsorption2DPesciSlowContour} shows the bulk concentration within the drop and the surface excess concentration along the interface for different dimensionless times $\hat{\tau}=\tau D / r_0^{2}$. It can be seen that a concentration gradient arises within the bulk, as surfactant adsorbs to the interface, resulting in diffusive transport of surfactant within the bulk to the interface. As the simple adsorption kinetics model given by \autoref{eq:linearKinetics} provides no desorption, the bulk gets depleted of surfactant, which accumulates completely at the interface. At time $\tau D / r_0^{2}=1$, we see the concentration fields already close to the respective equilibrium values $c_\mathrm{1,eq}=0$ and $\Gamma_\mathrm{\ell g,eq} = \nicefrac{1}{3}\,\SI{}{\mol\per\square\m}$.
For a quantitative evaluation of the method, it is useful to compare the simulation results with the analytical reference solution \citep{PesciDiss2019}, i.e.
\begin{equation}
\hat{c}_1(\hat{r},\hat{\tau})=\sum_{k=1}^\infty \Xi_k \frac{\sin (\Lambda_k \hat{r})}{\hat{r}} e^{-\Lambda_k^2 \hat{\tau}},
\label{eq:VerificationCurvedCBulkMixed}
\end{equation}
where $\hat{c}_1=c_1/c_{1,0}$ and $\hat{r}=r/r_0$ are dimensionless concentration and radial coordinate, respectively. From the boundary conditions for the bulk diffusion equation, the eigenvalues $\Lambda_k$ are given by \citep{PesciDiss2019}
\begin{equation}
\Lambda_k \cot \Lambda_k = 1-\frac{k_\mathrm{ad,H}r_0}{D}.
\end{equation}
Furthermore, the coefficients $\Xi_k$ result from the initial condition for the bulk concentration and are given by \citep{PesciDiss2019}
\begin{equation}
\Xi_k = \frac{\int_0^1 \hat{r} \sin(\Lambda_k \hat{r})\mathrm{d}\hat{r}}{\int_0^1 \sin^2(\Lambda_k \hat{r})\mathrm{d}\hat{r}}.
\label{eq:HenryIsotherm}
\end{equation}
For the comparison, the first $10^4$ terms of the series were evaluated. \autoref{fig:verificationAdsorption2DPesciSlowReference} shows the numerical solution alongside the analytical reference at the times corresponding to the contour plots shown in \autoref{fig:verificationAdsorption2DPesciSlowContour}. Excellent agreement between numerical and analytical solution can be observed from this comparison. Furthermore, another aspect of verifying the accuracy of the proposed numerical method is the conservative nature of total surfactant amount. During the adsorption process, the surfactant concentration within the bulk decreases, while the surface excess concentration increases. \autoref{fig:verificationAdsorption2DPesciSlowAmount} shows the species amount within the liquid bulk together with the species amount adsorbed to the interface over time. While the distribution of surfactant between interface and bulk changes during the adsorption process, the total surfactant amount remains constant.

\begin{figure}[htbp]
\centering
\begin{subfigure}[t]{1\textwidth}
\def\svgwidth{\linewidth}
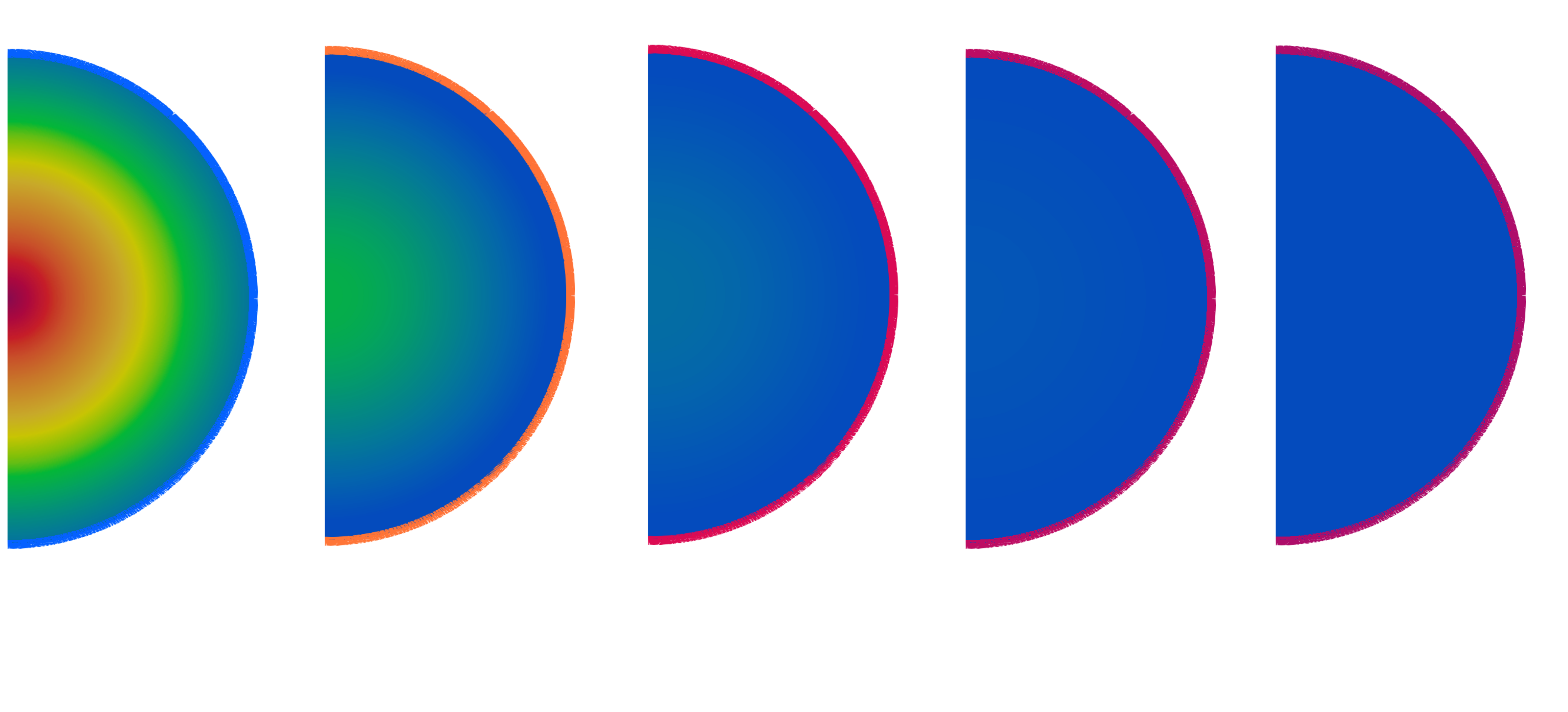
\caption{Bulk concentration and surface excess concentration from the numerical simulation.}
	\label{fig:verificationAdsorption2DPesciFastContour}
\end{subfigure}
\\
\vspace*{.5cm}
\begin{subfigure}[t]{.45\textwidth}
	\centering
	\includegraphics{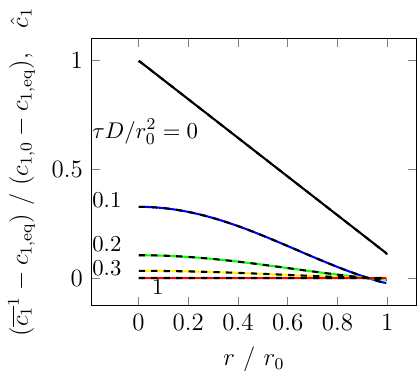}
	\caption{Bulk concentration from numerical solution (colored solid lines) in comparison to the analytical reference solution (black dashed lines) \citep{Antritter2022}.}
		\label{fig:verificationAdsorption2DPesciFastReference}
\end{subfigure}\hfill
\begin{subfigure}[t]{.45\textwidth}
	\centering
	\includegraphics{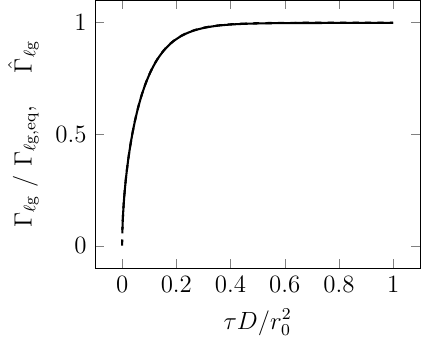}
	\caption{Surface excess concentration from numerical solution (solid line) in comparison to the analytical reference solution (dashed line) \citep{Antritter2022}.}
		\label{fig:verificationAdsorption2DPesciFastReferenceGamma}
\end{subfigure}

		\caption{Results from the verification case for diffusion limited adsorption from the bulk of a spherical drop to its interface. }
	\label{fig:verificationAdsorption2DPesciFast}
\end{figure}

\paragraph{Adsorption at the surface of a spherical drop: Diffusion limited}
This case considers adsorption that is not limited by the adsorption kinetics at the interface, but by diffusion through the bulk phase. As the previous case, it has been introduced by \citet{PesciDiss2019}. The geometry of this problem is identical to the one described in the previous section. Adsorption to the interface of a spherical drop with radius $r_0=\SI{1}{\m}$ from its bulk volume is studied. However, since there is no limiting adsorption kinetics now, surface excess concentration and subsurface bulk concentrations are in equilibrium. For the present test case, the Henry isotherm \citep[see e.g.][]{Chang1995}
\begin{equation}
\Gamma_\mathrm{\ell g}=K_\mathrm{H} c_1|_{r=r_0}
\end{equation}
is employed, where $K_\mathrm{H}=\SI{1}{\m}$ is the equilibrium constant. The initial conditions for the concentration fields are slightly adjusted.
The initial surface excess concentration is again set to $\Gamma_{\mathrm{\ell g}}(\tau{=}0)=\Gamma_{\mathrm{\ell g,0}}=0$. For the initial bulk concentration, a linear increase from the interface to the drop center according to $c_1(\tau{=}0,r)=c_{1,0}-\beta r$ is set, where $c_{1,0}=\SI{1}{\mol\per\cubic\m}$ and $\beta=\SI{0.8}{\mol\per\m\tothe{4}}$.
After introducing the dimensionless variables $\hat{c}_1=(c_1-c_\mathrm{1,eq})/(c_{1,0}-c_\mathrm{1,eq})$, $\hat{r}=r/r_0$ and $\hat{\tau}=\tau D / r_0^{2}$, where $c_\mathrm{1,eq}$ is the equilibrium concentration, the solution to this problem is again given by \autoref{eq:VerificationCurvedCBulkMixed} \citep{PesciDiss2019}. The eigenvalues $\Lambda_k$ are given by \citep{PesciDiss2019}
\begin{equation}
\Lambda_k \cot \Lambda_k = 1+\frac{K_\mathrm{H}}{r_0}\Lambda_k^2,
\end{equation}
accounting for the boundary conditions. 
The coefficients $\Xi_k$ are again calculated according to the initial conditions. For the present case they were computed using a least squares fit to the initial bulk concentration, while evaluating the first $10^4$ terms of the series. Furthermore, the dimensionless surface excess concentration $\hat{\Gamma}_\mathrm{\ell g}=\Gamma_\mathrm{\ell g}/\Gamma_\mathrm{\ell g,eq}$ is described by \citep{PesciDiss2019} 
\begin{equation}
\hat{\Gamma}_\mathrm{\ell g}=\frac{r_0 (c_{1,0}-c_\mathrm{1,eq})}{\Gamma_\mathrm{\ell g,eq} D}\sum_{k=1}^\infty \Xi_k \frac{\sin \Lambda_k - \Lambda_k\cos \Lambda_k}{\Lambda_k^2}\left(1-e^{-\Lambda_k^2 \hat{\tau}}\right).
\end{equation}
For the numerical solution, the linear sorption kinetics model
\begin{equation}
\frac{\partial\Gamma_\mathrm{\ell g}}{\partial\tau}=k_\mathrm{ad,H}c_1|_{r=r_0}-k_\mathrm{de,H}\Gamma_{\mathrm{\ell g}}
\end{equation}
is employed, which is consistent with the sorption isotherm in \autoref{eq:HenryIsotherm}. The adsorption and desorption rate coefficients $k_\mathrm{ad,H}=\SI{100}{\m\per\s}$ and $k_\mathrm{de,H}=\SI{100}{\per\s}$ are chosen sufficiently large, to quickly equilibrate surface excess and subsurface concentrations. Thereby, diffusion to the interface ($D=\SI{1}{\square\m\per\s}$) will be the limiting process for adsorption of surfactant to the interface in this test case.
An identical 2D axisymmetric computational mesh as in the previous case is employed, with a domain size of $\num{1.5}\, r_0\times \num{1.5}\, r_0$ discretized with $300\times 300$ cells.

\autoref{fig:verificationAdsorption2DPesciFast} shows the results from this test case. The evolution of bulk concentration and surface excess concentration from the numerical solution is shown in \autoref{fig:verificationAdsorption2DPesciFastContour}. It can be seen that, as surfactant adsorbs to the liquid-gas interface, the bulk concentration decreases, while the surface excess concentration increases.
Finally, both concentrations $\overline{c_1}^1$ and $\Gamma_\mathrm{\ell g}$ reach their respective equilibrium values $c_\mathrm{1,eq}=\SI{0.1}{\mol\per\cubic\m}$  and $\Gamma_\mathrm{\ell g,eq}=\SI{0.1}{\mol\per\square\m}$.
For a more quantitative comparison, \autoref{fig:verificationAdsorption2DPesciFastReference} shows these numerical results in comparison with the reference solution given above. Excellent agreement between the bulk concentration obtained using the method presented in this paper and the reference solution can be observed. Similarly, the corresponding surface excess concentration presented in \autoref{fig:verificationAdsorption2DPesciFastReferenceGamma} shows excellent agreement with the reference solution. Thus, overall, excellent agreement can be observed in this diffusion-limited adsorption case.

\subsubsection{Adsorption to a solid-fluid interface}
\label{sec:AdsorptionatSLinterface}
The verification of the proposed numerical methodology for surfactant adsorption at a solid-fluid interface is presented here. As mentioned before, the boundary condition at the solid-fluid interface presented in \autoref{eq:boundarycondition} is implemented as a mixed boundary condition within OpenFOAM (see \autoref{sec:BoundaryCondition}). 
The adsorption process follows the Langmuir-Hinshelwood kinetics \citep[see e.g.][]{Chang1995}, which can be regarded as a special case of the Langmuir-Freundlich kinetic model presented in \autoref{sec:AppendixFreundlich}. Thus, the adsorption rate at the solid-liquid interface is described by
\begin{equation}
\frac{\partial \Gamma_\mathrm{s\ell}}{\partial \tau}=k_\mathrm{ad,LH} c_1|_{x=0} \left(1-\frac{\Gamma_{\mathrm{s\ell}}}{\Gamma_\mathrm{s\ell,max}}\right) - k_\mathrm{de,LH} \left(\frac{\Gamma_{\mathrm{s\ell}}}{\Gamma_\mathrm{s\ell,max}}\right).
\label{eq:LangmuirFreundlichKinetics}
\end{equation}
\begin{figure}

\begin{subfigure}[t]{.43\textwidth}
\centering
\def\svgwidth{\linewidth}

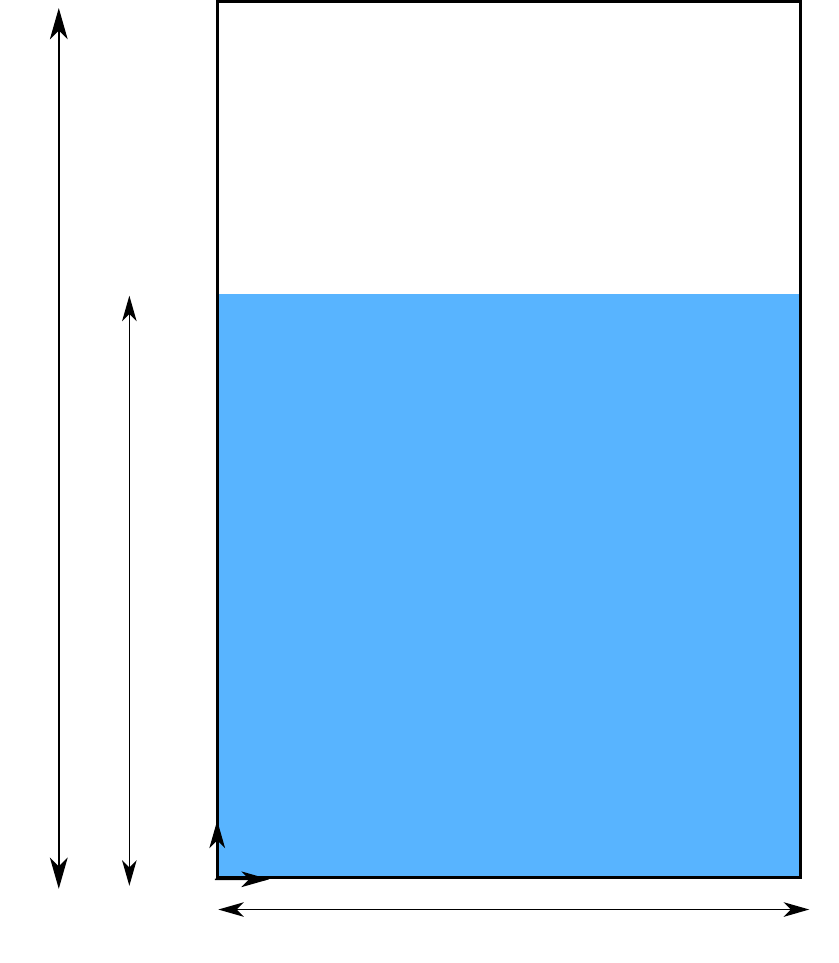
\caption{Computational domain and initial conditions}
\label{fig:domainWallAdsorption}
\end{subfigure}
\begin{subfigure}[t]{.55\textwidth}
\centering
\includegraphics{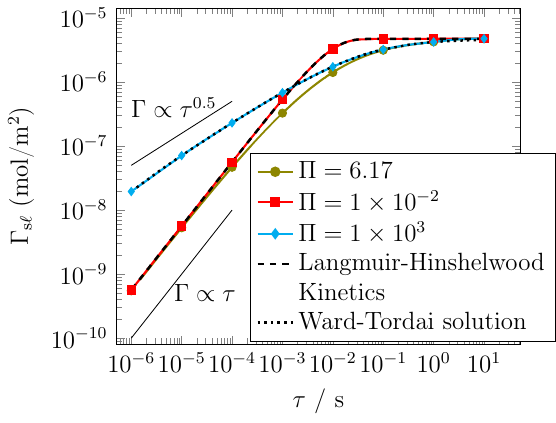}
	\caption{For various scenarios, the results obtained from the numerical simulations using Langmuir-Hinshelswood kinetics are compared with numerical (dotted) and analytical (dashed) reference solutions.   }
	\label{fig:WallAdsorptionGamma}
	\end{subfigure}

\begin{subfigure}{.95\textwidth}
\centering
\def\svgwidth{\linewidth}
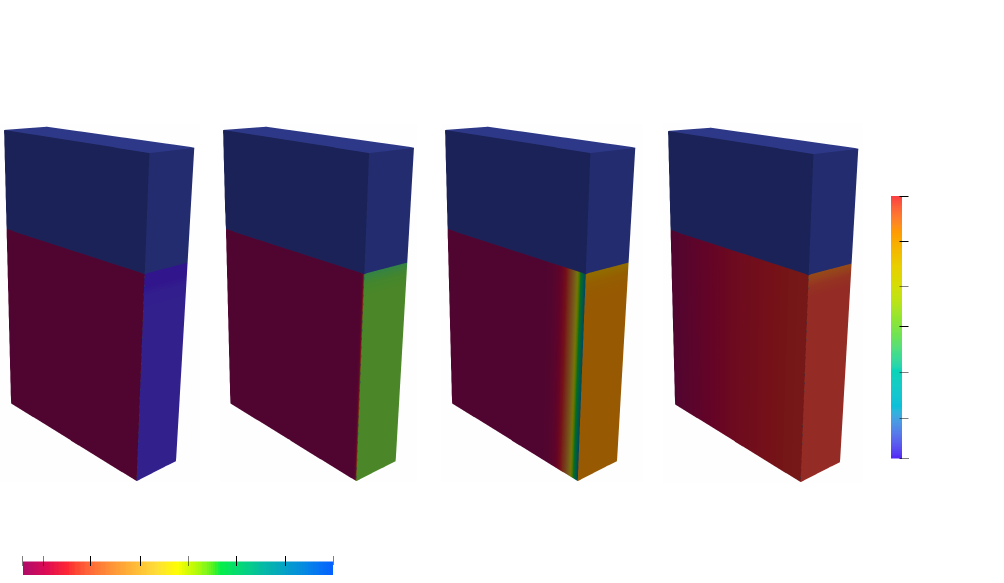
\par\bigskip
\par\bigskip
\caption{At various time instants, the snapshots show the bulk concentration and the surface excess concentration at the wall obtained from the numerical simulations, for the case with $\pi = 6.17$.}
\label{fig:WallAdsorptionFields}
\end{subfigure}

		\caption{Results from the verification case of adsorption at a solid-fluid interface.}
		\label{fig:WallAdsorption}
\end{figure}
The verification case is set up in analogy to a scenario presented in \citep{Antritter2019b}, where the adsorption at a liquid-gas interface is investigated for a one-dimensional problem. In the present study, a two-dimensional problem is considered, where a liquid-gas interface forms a $90^\circ$ contact angle with a solid substrate (see \autoref{fig:domainWallAdsorption}).  The material properties correspond to adsorption of heptanol at the interface of a heptanol-water solution and air \citep{Joos1989}. This adsorption process represents a mixed-kinetics case, i.e. an adsorption process, where both diffusion of surfactant within the bulk and the adsorption kinetics at the interface are important for the overall adsorption rate.  Diffusion coefficient, adsorption rate coefficients, and the maximum surface excess concentration are ${D=\SI{5.3e-10}{\square\m\per\s}}$, $k_\mathrm{ad,LH}=\SI{6e-4}{\m\per\s}$, $k_\mathrm{de,LH}=\SI{6.6e-3}{\mol\per\square\m\per\s}$, and $\Gamma_{\mathrm{s\ell ,max}}=\SI{6e-5}{\mol\per\square\m}$, respectively \citep{Joos1989}. An initial bulk and far-field concentration of $c_\mathrm{1,0}=c_{1,\infty}=\SI{0.941}{\mol\per\cubic\m}$ is chosen in analogy to \citep{Antritter2019b}.
In terms of dimensionless groups, those properties result in a ratio of kinetic to diffusion parameters $\Pi=\mathit{Da}\,{C}\,{G}^{-1}=\frac{\Gamma_\mathrm{s\ell,max} k_\mathrm{ad,LH}^2}{k_\mathrm{de,LH} D}=\SI{6.17}{}$ \citep{Antritter2022}. Therein, ${C}=c_{1,\infty}\,k_\mathrm{ad,LH}/k_\mathrm{de,LH}$ and ${G}=\Gamma_\mathrm{s\ell,eq}/\Gamma_\mathrm{s\ell ,max}$ are adsorption number and equilibrium surface coverage \citep{Ferri2000}, respectively. 

\autoref{fig:domainWallAdsorption} shows a schematic of the numerical domain and initial conditions. The size of the computational domain is $10h_\mathrm{ad}\times15h_\mathrm{ad}$, with \nicefrac{2}{3} of the domain filled with the liquid. Here, $h_\mathrm{ad} = \Gamma_\mathrm{s\ell,eq}/ c_{\mathrm{1,0}}$  is the characteristic length scale of a diffusion controlled adsorption process \citep{Ferri2000}. The numerical results are obtained on a computational grid with 300 $\times$ 30 cells. For this verification case, we consider adsorption only to the wall, with no adsorption to the liquid-gas interface. The wall is initially free of excess surfactant, $\Gamma_\mathrm{s}(\tau{=}0)=0$, while the bulk concentration in the liquid takes a homogeneous initial value $c_\mathrm{1}=c_\mathrm{1,0}$. In the gas phase, the initial bulk concentration is zero. The fluids are assumed to be at rest throughout the simulation. 
Similar to the cases presented in Section \ref{sec:AdsorptionDesorption}, the advection equation for the volume fraction field and the momentum equation are not solved. The boundary condition at the wall is implemented according to \autoref{eq:boundarycondition}, whereas, on the opposite side away from the solid-fluid interface, the bulk concentration of surfactant is fixed at $c_{1,\infty} = c_\mathrm{1,0}$. Therefore, the surfactant within the bulk diffuses towards the wall and gets adsorbed there. The temporal variation of the surface excess concentration, $\Gamma_\mathrm{s\ell}$ for this case with $\Pi = 6.17$, is shown in \autoref{fig:WallAdsorptionGamma}. Furthermore, \autoref{fig:WallAdsorptionFields} shows the snapshots at various instants of time, highlighting the variation of surfactant concentration within the bulk along with the surface excess concentration at the wall. First, it can be observed that  up to about $10^{-4}$s, the concentration of surfactant within the bulk is almost constant, highlighting that the process is adsorption-limited, which is further indicated by the slope $\Gamma_\mathrm{s\ell} \propto \tau$ (see \autoref{fig:WallAdsorptionGamma}). Followed by this, a gradient in concentration of surfactant near the solid-fluid interface can be observed, indicating a transition to diffusion limited regime. This mixed kinetics for the used exemplary properties of heptanol water solution is expected. As $\Gamma_{\mathrm{s\ell}}$ approaches $\Gamma_{\mathrm{s\ell,eq} }$, the surfactant is mostly replenished as can be seen from a more uniform distribution of bulk concentration at a time instant of 1s.

In addition to the mixed-kinetics case, additional cases corresponding to the slow adsorption (implying adsorption kinetics limited) regime, as well as the fast adsorption (implying diffusion limited) regime are considered. For the case with adsorption at the solid-fluid interface being the rate determining step, the diffusion coefficient is increased to ${D=\SI{32.72e-8}{\square\m\per\s}}$, while keeping the kinetic adsorption coefficients $k_{\mathrm{ad,LH}}$ and $k_{\mathrm{de,LH}}$ constant, resulting in $\Pi= 1\times10^{-2}$. Similarly, for the case when diffusion within the bulk is the rate determining step, the kinetic adsorption coefficients are increased to $k_\mathrm{ad,LH}=\SI{9.62e-4}{\m\per\s}$ and $k_\mathrm{de,LH}=\SI{1.05}{\mol\per\square\m\per\s}$, maintaining their ratio and therefore the equilibrium state, while keeping the diffusion coefficient constant at ${D=\SI{5.3e-10}{\square\m\per\s}}$. This then results in $\Pi = 1\times10^{3}$. Similar to the case of mixed kinetics, the numerical results for the adsorption limited case are obtained on a computational grid with 300 $\times$ 30 cells. For the diffusion limited case, the numerical results are obtained on a mesh graded by a factor of 100 in the $x$-direction, the smallest cells being near to the solid-fluid interface, where the concentration gradients are higher in the early stages of the simulation. 

For a quantitative evaluation of the results for these extreme cases of sorption processes, we again draw a comparison with the respective reference solutions. Due to the similar setup and parameter choice, the reference solutions presented in \citep{Antritter2019b} for the adsorption at the liquid-gas interface can also be used here for the adsorption at the solid-fluid interface. For the kinetics-limited case, the reference solution is given by
\begin{equation}
\label{eq:LH-kinetics-ODE}
\Gamma_\mathrm{s\ell}(\tau) = \frac{\Gamma_\mathrm{s\ell,max} k_\mathrm{ad,LH} c_{1,\infty}}{k_\mathrm{ad,LH}c_{1,\infty} + k_\mathrm{de,LH}} \left\lbrace 1 - \exp{ \left[ -\left(\frac{k_\mathrm{ad,LH}c_{1,\infty} + k_\mathrm{de,LH}}{\Gamma_\mathrm{s\ell,max}}\right) \tau \right] } \right\rbrace,
\end{equation}
obtained from the solution of the ordinary differential equation given by the Langmuir-Hinshelwood kinetics together with a constant subsurface bulk concentration of $c_\mathrm{1,\infty}$ \citep{Antritter2019b}. For the diffusion limited case, the adsorption process can be described by the Ward-Tordai \citep{Ward1946} equation,
\begin{equation}
\label{eq:WardTordai}%
\Gamma_\mathrm{s\ell}(\tau) = 2 \sqrt{ \frac{D}{\pi} } \left( c_{1,\infty} \sqrt{\tau} - \int_0^{\sqrt{\tau}} c_1|_{x=0} (\hat{\tau}) \mathrm{d}\sqrt{\tau-\hat{\tau}} \right),
\end{equation}
while assuming an equilibrium of the subsurface bulk concentration and surface excess concentration according to the Langmuir isotherm. For this case, a numerical reference solution is obtained using the C++ code provided by \citet{Li2010}.

\autoref{fig:WallAdsorptionGamma} shows the surface excess concentration from the numerical simulations (for $\pi = 1 \times 10^{-2}$ and $1\times 10^{3}$) in comparison with the reference solutions (dashed line and dotted line, respectively).
Excellent agreement between the numerical model and the reference solutions is observed over a large time interval, for both of the extreme cases of adsorption and diffusion limited sorption kinetics. Comparing the previously presented case of mixed-kinetics with these results, the increase in surface excess concentration initially follows the reference solution of the adsorption limited case. At later times, diffusion to the interface becomes the limiting factor. Thus, the surface excess concentration shows a slower increase compared to the kinetics-limited case, until finally following the reference solution for the diffusion-controlled limit.

\subsection{Marangoni Flow}
\label{sec:MarangoniFlow}
The test cases shown in \autoref{sec:VerificationConvection} focused on the coupling of adsorption processes and species transport to the flow via the velocity field. However, for surface active substances, the coupling in the opposite direction is relevant as well. As the surface tension depends on the local surface excess concentration of surfactant, Marangoni stresses may arise, which enter the momentum balance and affect the flow. In order to verify our method for this effect, the flow within and around a viscous drop with a fixed, prescribed surfactant distribution on its surface is studied. This case has previously been studied for sharp \citep{Muradoglu2008} and diffuse \citep{Teigen2011} interface methods. The drop is assumed to be of spherical shape with a radius $r_0$ and is in a zero gravity environment. The surfactant is distributed along the drop surface such that the surface tension varies according to
\begin{equation}
\sigma_\mathrm{\ell g}(z)=\sigma_\mathrm{\ell g,0} \left(\frac{1}{2}-\frac{z}{15\,r_0}\right),
\label{eq:sigmaMarangoniValidation}
\end{equation}
with the linear surface tension model
\begin{equation}
\sigma_\mathrm{\ell g} = \sigma_\mathrm{\ell g,0} \left(1-\frac{\Gamma_\mathrm{\ell g}}{\Gamma_\mathrm{\ell g,max}}\right)
\label{eq:linearSurfaceTension}
\end{equation}
employed.

\begin{figure}[h!]

\pgfplotstablesort[sort key=Points:2]{\sorted}{plotData/VerificationValidation/Marangoni_TeigenGrad/DS2/DropContour0.csv}
\begin{subfigure}[t]{.45\textwidth}
	\centering
	\includegraphics{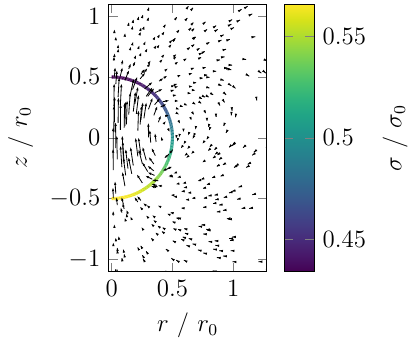}
	\caption{Velocity and surface tension fields}
	\label{fig:VerificationMarangoniUSigma}
\end{subfigure}
\hfill
\begin{subfigure}[t]{.45\textwidth}
\centering
		\includegraphics{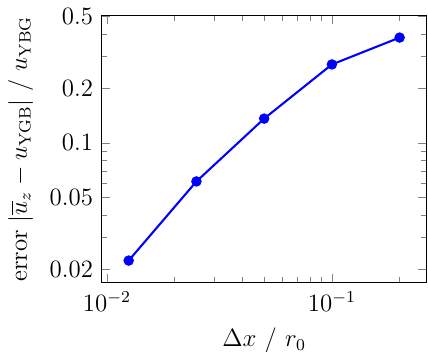}
	\caption{Error in mean bubble velocity compared to that predicted by \citet{Young1959}.}
		\label{fig:ErrorVerificationMarangoni}
\end{subfigure}
\caption{Results from the verification case of a rising bubble due to Marangoni stresses \citep{Antritter2022}.}
\end{figure}

In the above equations, $z$ denotes the coordinate along the axis of symmetry of the problem, with $z=0$ marking the drop center. This is identical to the profile employed by \citet{Teigen2011}, with the origin of the coordinate system shifted to the drop center. Similarly, all relevant parameters $\rho_1=\rho_2=\SI{0.2}{\kg\per\cubic\m}$, $\mu_1=\mu_2=\SI{0.1}{\Pa\s}$, $\sigma_\mathrm{\ell g,0}=\SI{1.0}{\N\per\m}$ and $r_0=\SI{0.5}{\m}$ are chosen in accordance with the test case presented by \citet{Teigen2011}. The surfactant distribution as well as the drop shape are assumed fixed for this test case, i.e. they do not vary with time. Then, the analytical solution derived by \citet{Young1959} for drop motion due to a surface tension gradient can be used as reference. For the present case, the expected terminal velocity of the drop is then \citep[cf.][]{Teigen2011}
\begin{equation}
u_\mathrm{YGB}=\frac{2\sigma_\mathrm{\ell g,0}}{15(6\mu_1+9\mu_2)}=\frac{2}{225}\frac{\sigma_\mathrm{\ell g,0}}{\mu_1}=\frac{4}{45}\SI{}{\m\per\s}=\SI[parse-numbers=false]{0.0\overline{8}}{\m\per\s}.
\label{eq:uYGB}
\end{equation}

For the numerical solution, a computational domain with a size $16\,r_0 \times 8\,r_0$ in axial and radial dimension, respectively, is used. The domain thus makes use of the axial symmetry of the problem. Atmospheric conditions\footnote{A constant total pressure is set. For the velocity field, a homogeneous Neumann condition is set if the flow is directed out of the domain. For an inflow, a homogeneous Neumann condition is set for the boundary normal component, while the tangential component is set to zero.} are set along the boundaries. The mesh is refined in two steps from the boundary towards the drop. The influence of the spatial discretization is studied by using grids with mesh sizes varying from $\Delta x = \num{0.2}\,r_0$ at the drop for the coarsest grid to $\Delta x = \num{1.25e-2}\,r_0$ for the finest grid. The evaluation method of $\Gamma_\mathrm{\ell g}$ from $c_\mathrm{I}$ and $\alpha$ is included in this verification case by evaluating the local surface tension from equations \autoref{eq:GammaFromS} and \autoref{eq:linearSurfaceTension}.

The velocity field from the numerical solution obtained on the finest grid is shown in \autoref{fig:VerificationMarangoniUSigma} alongside the surface tension evaluated from the prescribed surfactant distribution. Due to the lower surface tension at the top, a pressure gradient within the drop arises from the difference in Laplace pressures. The resulting flow from the bottom of the drop to its top, resulting also in the drop motion, can be observed in the velocity field. However, counteracting this flow is the recirculation along the drop's interface driven by Marangoni stresses due to the surfactant concentration gradient. For a quantitative comparison to the reference solution given in \autoref{eq:uYGB}, the drop velocity along the axial direction is calculated from the numerical results according to 
\begin{equation}
\overline{u}_z=\frac{\sum_{i=1}^N u_{i,z}\alpha_i V_i}{\sum_{i=1}^N \alpha_i V_i}.
\end{equation}
Therein, $u_{i,z}$ and $\alpha_i$ represent the axial velocity and volume fraction within the computational cell $i$ with volume $V_i$. The relative error of the predicted drop velocity, compared to the reference solution, is shown in \autoref{fig:ErrorVerificationMarangoni} as a function of grid resolution. The figure shows that the error decreases with mesh refinement and reaches a level of a few percent on the finer considered grids. Thus, the presented method, including the evaluation of the local surface excess concentration, accurately captures solutal Marangoni flow.

\section{Conclusion and Outlook}
\label{sec:Conclusion}
Allowing to account for surface active substances further improves the versatility of the algebraic volume of fluid method. In the present work, this is achieved by a two-field formulation, where the surfactant concentration at the liquid-gas interface and within the bulk are tracked separately. Specifically, the surface excess concentration at a liquid-gas interface is tracked using a volumetric concentration, taking advantage of the slightly diffuse interface region. Interestingly, this approach, utilizing the diffuse nature of the interface region, was employed before in the context of C-LS and phase field methods, but not within the algebraic VOF method. Furthermore, considering the solubility of surfactant within the bulk, the numerical framework can handle adsorption and desorption at the liquid-gas interface by coupling the two transport equations using source terms. Adsorption to the solid substrate from the liquid bulk is treated by the presented method via an appropriate boundary condition at the solid and an additional species balance for the solid wall. This therefore enables the method to be also applied to wetting phenomena, where adsorption of surfactant to the substrate may play a role. Even though the method discussed here is in the context of liquid-gas systems potentially in contact with a solid, the method is not limited to such systems, and can also be applied to liquid-liquid two phase systems. The method has been applied to several verification cases in order to demonstrate its applicability and accuracy with respect to surfactant transport and adsorption processes, as well as Marangoni flow.

In future works, we intend to apply the presented method to the impact and spreading of surfactant laden droplets with sizes on the micrometer scale on a solid substrate. For such cases, the adsorption of surfactant to the substrate is a point of special interest. Such phenomena are relevant to inkjet printing and other, similar coating processes. Due to the short length and time scales for the hydrodynamics of such processes, the diffusive transport towards the interface can be resolved using the presented method. In order to apply our method also to larger scales, subgrid-scale modeling along the lines of the work of \citet{weiner2017advanced} and \citet{Pesci2018} may be introduced in analogy to the current implementation of the sorption kinetics in the method presented here.

\section*{Acknowledgements}
We gratefully acknowledge the financial support by the German Research Foundation (DFG) - Project-ID 265191195 within the Collaborative Research Centre 1194 "Interaction of Transport and Wetting Processes", sub-projects T01, B02 and Z-INF. Furthermore, we gratefully acknowledge the financial support by Heidelberger Druckmaschinen AG for the associated project preceding T01. Calculations for this research were conducted on the Lichtenberg high performance computer of the TU Darmstadt.

\bibliography{literature}

\appendix
\newpage
\section{Linearization of Langmuir-Freundlich kinetics}
\label{sec:AppendixFreundlich}
Non-linear sorption models, such as the Langmuir-Freundlich kinetic model, can be considered in the framework presented within this article by linearization around the current surface excess and sub-surface concentrations. The Langmuir-Freundlich kinetic model describes the adsorption rate as \citep{Chan2012}
\begin{equation}
\frac{\partial \Gamma}{\partial \tau}=k_\mathrm{ad,LF} c_\mathrm{1,I} \left(1-\frac{\Gamma}{\Gamma_\mathrm{max}}\right)^{\nicefrac{1}{n}} - k_\mathrm{de,LF} \left(\frac{\Gamma}{\Gamma_\mathrm{max}}\right)^{\nicefrac{1}{n}},
\label{eq:LangmuirFreundlichKinetics}
\end{equation}
where $k_\mathrm{ad,LF}$ and $k_\mathrm{de,LF}$ are the adsorption and desorption rate coefficients and $\Gamma_\mathrm{max}$ is the maximum surface excess concentration. The exponent $n$ describes cooperativity of adsorbing surfactant molecules \citep{Guzman2011}. The above kinetics are consistent with the Langmuir-Freundlich isotherm, which is also known as Sips isotherm \citep{Sips1948}
\begin{equation}
\frac{\Gamma}{\Gamma_\mathrm{max}}=\frac{(K_\mathrm{LF} c)^n}{1+(K_\mathrm{LF} c)^n},
\label{eq:LangmuirFreundlichIsotherm}
\end{equation}
describing the equilibrium of concentrations (i.e. when $\partial \Gamma / \partial \tau = 0$). The equilibrium coefficient therein is related to the rate coefficients by $K_\mathrm{LF}= k_\mathrm{ad} / k_\mathrm{de}$.
It is to be noted that, for the sake of generality, the above equations and the following discussion on the linearization is presented by representing the surface excess concentration as $\Gamma$. This methodology and the sorption model can be applied to either $\Gamma_{\mathrm{\ell g}}$ or $\Gamma_{\mathrm{s\ell}}$.

In order to arrive at the form as presented in \autoref{eq:interfaceSourceTerms} for this non-linear sorption kinetics model, \autoref{eq:LangmuirFreundlichKinetics} can be linearized around the current surface excess concentration $\Gamma = \Gamma_0$ using Taylor series expansion. This linearized form then allows semi-implicit treatment of the source terms for the surface excess concentration.
Using only the leading order-terms of the Taylor series up to the linear term, \autoref{eq:LangmuirFreundlichKinetics} can be approximated by
\begin{align}
\left.\frac{\partial \Gamma}{\partial \tau}\right\vert_{\Gamma_0}
\approx & \mathbin{\hphantom{-}} k_\mathrm{ad,LF} c_\mathrm{1,I} \left(1-\frac{\Gamma_0}{\Gamma_\mathrm{max}}\right)^{\nicefrac{1}{n}} - k_\mathrm{de,LF} \left(\frac{\Gamma_0}{\Gamma_\mathrm{max}}\right)^{\nicefrac{1}{n}} \nonumber \\
& -\left[ k_\mathrm{ad,LF} c_\mathrm{1,I} \left(1-\frac{\Gamma_0}{\Gamma_\mathrm{max}}\right)^{\frac{1-n}{n}} + k_\mathrm{de,LF} \left(\frac{\Gamma_0}{\Gamma_\mathrm{max}}\right)^{\frac{1-n}{n}}\right]\ \frac{\Gamma-\Gamma_0}{n\, \Gamma_\mathrm{max}}.
\label{eq:LangmuirFreundlichHinshelwoodLinearized}
\end{align}
Rearranging into constant and linear terms then results in the linearized sorption model
\begin{align}
\left.\frac{\partial \Gamma}{\partial \tau}\right\vert_{\Gamma_0}
\approx & \hphantom{-} \underbrace{k_\mathrm{ad,LF} c_\mathrm{1,I} \left(1-\frac{\Gamma_0}{\Gamma_\mathrm{max}}\right)^{\frac{1-n}{n}} \left[\left(1-\frac{\Gamma_0}{\Gamma_\mathrm{max}} \right) + \frac{\Gamma_0}{n\,\Gamma_\mathrm{max}}\right]
\ +\ k_\mathrm{de,LF} \frac{1-n}{n}\left(\frac{\Gamma_0}{\Gamma_\mathrm{max}}\right)^{\nicefrac{1}{n}}}_A \nonumber \\
& \underbrace{-\frac{1}{n\, \Gamma_\mathrm{max}}\ \left[ k_\mathrm{ad,LF} c_\mathrm{1,I} \left(1-\frac{\Gamma_0}{\Gamma_\mathrm{max}}\right)^{\frac{1-n}{n}} + k_\mathrm{de,LF} \left(\frac{\Gamma_0}{\Gamma_\mathrm{max}}\right)^{\frac{1-n}{n}}\right]}_B\ \Gamma.
\label{eq:LangmuirFreundlichHinshelwoodLinRearranged}
\end{align}
In this linearized form, the sorption model can already be used to describe the adsorption process to the solid substrate in \autoref{eq:boundarycondition} and \autoref{eq:balanceGammasld} as
\begin{equation}
j_\mathrm{s\ell}=A + B \Gamma_\mathrm{s\ell}.
\end{equation}
For the adsorption at the liquid-gas interface, an additional step is necessary. In analogy to the work of \citet{Lakshmanan2010}, the linearized kinetic model can be transferred into the volumetric form
\begin{equation}
j_\mathrm{I}\, =\, A\, \delta_\mathrm{I} + B\, c_\mathrm{I}
\label{eq:interfaceSourceTermsappendix}
\end{equation}
by weighting with $\delta_\mathrm{I}=||\nabla\alpha||_2$. For $n=1$, the expression reduces to the volume-averaged Langmuir-Hinshelwood kinetics
\begin{equation}
j_\mathrm{I}=k_\mathrm{ad} c_\mathrm{1,I}\left(\delta_\mathrm{I}-\frac{c_\mathrm{I}}{\Gamma_\mathrm{max}}\right)-k_\mathrm{de} c_\mathrm{I},
\label{eq:jI}
\end{equation}
as it has previously been presented in \citep{Antritter2019b}, similar to the work of \citet{Lakshmanan2010}.\\

\paragraph{Surface equation of state}
Adsorption isotherm and surface tension are related through Gibb's adsorption equation
\begin{equation}
\Gamma = -\frac{1}{\overline{R}T}\left(\frac{\partial \sigma}{\partial \ln c}\right)_{T,p},
\end{equation}
where $\overline{R}$ is the ideal gas constant, $T$ is the temperature, and $\sigma$ is the surface tension or surface energy of the respective interface.
The reduction in surface tension due to the adsorption of surfactant for the Langmuir-Freundlich isotherm, \autoref{eq:LangmuirFreundlichIsotherm}, is therefore given by
\begin{equation}
\sigma_\mathrm{0} - \sigma = \overline{R}T\Gamma_\mathrm{max} n^{-1} \ln \left(1+(K_\mathrm{LF}c)^n\right).
\end{equation}
Therein, $\sigma_\mathrm{0}$ is the surface tension without any adsorbed excess surfactant. Using \autoref{eq:LangmuirFreundlichIsotherm}, one arrives at the surface equation of state
\begin{equation}
\sigma_\mathrm{0} - \sigma = -\overline{R}T\Gamma_\mathrm{max} n^{-1} \ln \left(1-\frac{\Gamma}{\Gamma_\mathrm{max}}\right),
\end{equation}
consistent with the Langmuir-Freundlich isotherm, describing the local surface tension or surface energy as a function of the local surface excess concentration.

\newpage
\section{Discretization schemes}
\label{sec:schemes}
Throughout this paper we employ discretization schemes provided by the OpenFOAM library. Boundedness of the variables is taken into account in the selection of schemes. \autoref{tab:schemes} lists the schemes employed throughout this work.

\begin{table}[htbp]
\caption{Discretization schemes used throughout this paper, following the naming in OpenFOAM (see e.g. \citep{OpenFOAM2016,Greenshields2016}).}
\label{tab:schemes}
\begin{longtable}{@{\extracolsep{\fill}} p{0.2\textwidth} p{0.3\textwidth} p{0.3\textwidth}}
\toprule
\textbf{Operator} & \textbf{Term} & \textbf{Scheme} \\
\midrule
\endfirsthead
$\partial\varphi/\partial\tau^{(*)}$ 	& default & Euler \\
\midrule
$\nabla\varphi$ 				& default & Gauss linear \\
\midrule
$\nabla\cdot\varphi$		 	& $\mathbf{u}\alpha$ & Gauss SFCD \\
								& $\mathbf{u}_\mathrm{r}(1-\alpha)\alpha$ & Gauss SFCD \\
							 	& $\rho_\mathrm{m}\mathbf{u}\otimes\mathbf{u}$ & Gauss SFCD \\		
							 	& $\mathbf{u} c_\mathrm{B}$ & Gauss SFCD \\
							 	& $\mathbf{u}_\mathrm{r}(1-\alpha)c_\mathrm{B}$ & Gauss SFCD \\
							 	& $\frac{D_1 \nabla\alpha}{\alpha+\varsigma_\alpha}c_\mathrm{B}$ & Gauss linear \\
							 	& $\mathbf{u} c_\mathrm{I}$ & Gauss vanLeer \\
							 	& $\mathbf{u}_\mathrm{r}w_2(\alpha)c_\mathrm{I}$ & Gauss vanLeer \\
							 	& $\mathbf{u}_\mathrm{c,\delta}w_1(\alpha)c_\mathrm{I,S}$ & Gauss SFCD \\
							 	& $\mathbf{u}_\mathrm{c,\delta}w_1(\alpha)\delta_\mathrm{I,S}$ & Gauss SFCD \\
							 	& $\mathbf{u}_\mathrm{r}(1-\alpha)j_\mathrm{B}$ & Gauss SFCD \\
							 	& $\frac{D_{j,\mathrm{B}} \nabla\alpha}{\alpha+\varsigma_\alpha}j_\mathrm{B}$ & Gauss linear \\
							 	\midrule
interpolation					& default	& linear \\
\midrule
$\nabla\cdot\nabla \varphi$		& default	& Gauss linear limited 0.3\\
\midrule
$\mathbf{n}_f\cdot\nabla\varphi$		& default	& limited 0.3 \\
\bottomrule
\end{longtable}
\end{table}

\newpage
\section{Boundary condition at a solid-fluid interface}
\label{sec:BoundaryCondition}
The mixed boundary condition in OpenFOAM calculates the face value of any field $\phi$ at a boundary as \citep{Le2012}
\begin{equation}
\phi_{f}=w \phi_\mathrm{ref}+(1-w)\left(\phi_{c}-\Delta x\nabla \phi_\mathrm{ref} \right),
\label{eq:openfoamBC}
\end{equation}
where $\phi_{f}$ is the face value of the variable, $\phi_{c}$ is the value at the cell center, $\phi_\mathrm{ref}$ is the reference value, $\Delta x$ is the face-to-cell distance, and $w$ is the value fraction. 

The boundary condition for the case of adsorption and desorption at a solid-fluid interface is
\begin{equation}
\mathbf{n}_\mathrm{w}\cdot D\nabla c_\mathrm{B} - \frac{c_\mathrm{B}}{\alpha+\varsigma_\alpha}\, \mathbf{n}_\mathrm{w}\cdot D\nabla\alpha = j_\mathrm{s}.
\label{eq:BC}
\end{equation}
Following a similar convention as that of \autoref{eq:openfoamBC}, \autoref{eq:BC} can be rewritten as
\begin{equation}
\frac{c_{\mathrm{B},f}-c_{\mathrm{B},c}}{\Delta x}+ \frac{c_{f}}{\alpha_{f}+\varsigma_\alpha }
\operatorname{snGrad}(\alpha)_{f} = \frac{\alpha_{f}}{D} j_{\mathrm{s}},
 \end{equation}
where $\operatorname{snGrad}$ represents the discrete surface normal gradient.
After rearranging to
 \begin{equation}
   \frac{c_{\mathrm{B},c}}{\Delta x}=c_{\mathrm{B},f}\left(\frac{1}{\Delta x}-\frac{1}{\alpha_{f}+\varsigma_\alpha} \operatorname{snGrad}(\alpha)_{f}\right) +\frac{\alpha_{f}}{D} j_{\mathrm{s}}
 \end{equation}
and further to
 \begin{equation}
  c_{\mathrm{B},f}=\left(\frac{c_{\mathrm{B},c}}{\Delta x}-\frac{\alpha_{f}}{D} j_{\mathrm{s}} \right) \frac{1}{\left(\frac{1}{\Delta x}-\frac{1}{\alpha_{f}+\varsigma_\alpha}\operatorname{snGrad}(\alpha)_{f} \right) },
 \end{equation}
one arrives at
\begin{equation}
c_{\mathrm{B},f} =\left(c_{\mathrm{B},c}-\frac{\alpha_{f}}{D} \Delta j_{\mathrm{s}} \right) \frac{1}{\left(1-\frac{\Delta x}{\alpha_{f}+\varsigma_\alpha} \operatorname{snGrad}(\alpha)_{f}\right)}.
\label{eq:cBBoundary}
\end{equation}
From comparison of coefficients between \autoref{eq:openfoamBC} and \autoref{eq:cBBoundary}, the boundary condition at a solid-fluid interface can, therefore, be expressed as
\begin{equation}
 c_{\mathrm{B},f}= (1-w)\left(c_{\mathrm{B},c}-\Delta x (\nabla c_{\mathrm{B}})_\mathrm{ref} \right),
\end{equation}
where
\begin{equation}
(\nabla c_{\mathrm{B}})_\mathrm{ref} = \frac{\alpha_{f}}{D} j_{\mathrm{s}} 
\end{equation}
and
\begin{equation}
1-w= \frac{1}{\left(1-\frac{\Delta x}{\alpha_{f}+\varsigma_\alpha} \operatorname{snGrad}(\alpha)_{f}\right)}.
\end{equation}
For the reference value, the comparison of coefficients shows
\begin{equation}
c_\mathrm{\mathrm{B},ref}=0.
\end{equation}

\end{document}